\documentclass[10pt,letterpaper]{article}
\usepackage{fullpage}

\usepackage{color}
\usepackage[table]{xcolor}
\usepackage[normalem]{ulem}

\usepackage{amsmath,amssymb}

\usepackage[utf8x]{inputenc}

\usepackage{textcomp,marvosym}

\usepackage{cite}

\usepackage{nameref,hyperref}

\usepackage{microtype}
\DisableLigatures[f]{encoding = *, family = * }

\usepackage[table]{xcolor}

\usepackage{array}

\usepackage{subcaption}

\newcommand{\eat}[1]{}

\usepackage[aboveskip=1pt,labelfont=bf,labelsep=period,justification=raggedright,singlelinecheck=off]{caption}

\usepackage{graphicx}
\usepackage{epstopdf}

\graphicspath{{./FIGS/}}
\newcommand{\figref}[1]{Fig.~\ref{#1}}
\newcommand{\specialheads}[1]{\noindent{\bf #1:\\}}

\DeclareMathOperator*{\argmin}{\arg\,\min}
\def\x{\mathbf{x}}

\usepackage{authblk}

\title{Rapid prediction of crucial hotspot interactions for icosahedral viral capsid
self-assembly by energy landscape atlasing validated by mutagenesis}

\author[1]{Ruijin Wu}
\author[1]{Rahul Prabhu}
\author[1]{Aysegul Ozkan}
\author[1]{Meera Sitharam$^*$}
\affil[1]{ Department of Computer and Information Science and Engineering,
University of Florida, Gainesville, Florida, United States of America}
\affil[ ]{*Email: sitharam@cise.ufl.edu}

\bigskip

\begin{document}
\maketitle
\section*{Abstract}
Icosahedral viruses are under a micrometer in diameter, their infectious genome
encapsulated by a shell assembled by a multiscale process, starting from an
integer multiple of 60 viral capsid or coat protein (VP) monomers.

We predict and validate inter-atomic hotspot interactions between VP monomers
that are important for the assembly of 3 types of icosahedral viral capsids:
Adeno Associated Virus serotype 2 (AAV2) and Minute Virus of Mice (MVM), both
$T=1$ single stranded DNA viruses, and Bromo Mosaic Virus (BMV), a $T=3$ single
stranded RNA virus. Experimental validation is by in-vitro, site-directed
mutagenesis data found in literature. 

We combine ab-initio predictions at two scales: at the \emph{interface-scale},
we predict the importance (\emph{cruciality}) of an interaction for successful
subassembly across each interface between symmetry-related VP monomers; and at
the \emph{capsid-scale,} we predict the cruciality of an interface for
successful capsid assembly.

At the interface-scale, we measure cruciality by changes in the capsid
free-energy landscape \emph{partition function} when an interaction is removed.
The partition function computation uses \emph{atlases} of interface subassembly
landscapes, rapidly generated by a novel geometric method and curated
opensource software EASAL (efficient atlasing and search of assembly
landscapes). At the capsid-scale, cruciality of an interface for successful
assembly of the capsid is based on combinatorial entropy.

Our study goes all the way from resource-light, multiscale computational
predictions of crucial hotspot inter-atomic interactions to 
validation using data on site-directed mutagenesis' effect on capsid
assembly. By reliably and rapidly narrowing down target interactions, (no
more than 1.5 hours per interface on a laptop with Intel Core
i5-2500K @ 3.2 Ghz CPU and 8GB of RAM) our predictions can inform and reduce
time-consuming in-vitro and in-vivo experiments, or more
computationally intensive in-silico analyses.
 

\section*{Author summary}
Viruses, found in all classes of living orgaisms, can be beneficial as well as
harmful to their hosts. Understanding their mechanism of assembly is critical
to understanding how we can inhibit or enhance their life cycle process. 


Icosahedral viral capsids, as elucidated by Caspar and Klug
\cite{caspar1962physical}, are self-assembled from nearly identical viral
capsid or coat-protein (VP) monomers spontaneously and rapidly, with high
efficacy and accuracy, a process sometimes facilitated by other biomolecules.
Understanding virus assembly requires identifying crucial VP-VP hotspot
interactions whose removal would disrupt the process. 
 
We combine a novel geometric method for rapidly atlasing free energy landscapes
with a symmetry-based combinatorial method to give a two-scale prediction of
hotspot interactions. We validate the predictions for 3 types of viruses, using
in-vitro, site-directed mutagenesis' disruptive effects on capsid assembly,
found in literature, noting that the biophysical assays for AAV2 were carried
out by the Mavis Agbandje-Mckenna's lab \cite{UFDC000005707} contemporaneously
with the development of our computational model and prediction. Our predictions
are reproducible using our curated opensource software EASAL (efficient
atlasing and search of assembly landscapes) \cite{maineasal, Ozkan:toms}.
 
To the best of our knowledge, prevailing methods for statistical mechanical
prediction of hotspot interactions use a single scale, are knowledge-based, are
computationally intensive, or have not been validated by in-vitro site directed
mutagenesis results.

\section{Introduction}
\label{sec:introduction}

Viruses can be pathogenic or non-pathogenic, rod-like or icosahedral, enveloped
or non-enveloped. Pathogenic viruses are detrimental to their host and
significant research is focused on their prevention, by disrupting crucial
steps in their life cycle \cite{schlicksup2018hepatitis}. Virus capsid assembly
is a critical step in the generation of infectious virus particles during their
replicative life cycle. Understanding assembly processes in the viral life
cycle illuminates the pathophysiology of infectious diseases, and allows us to
target assembly processes with drugs. Improving assembly of non-pathogenic
viruses can be utilized for certain beneficial applications, for example cancer
treatment with oncolytic viruses, cell and gene therapy applications, and for
vaccine production \cite{GoodVirus}.

Icosahedral viral capsids assembled from almost identical viral capsid or coat
protein (VP) monomers were elucidated by Caspar and Klug
\cite{caspar1962physical}. The number of VP monomers is some multiple (called
the $T$ number) of 60. At each inter-monomeric, symmetry-related interface, the
assembly is a nanoscale process influenced by inter-atomic interactions, while
the entire capsid can be between 10's to 100's of nanometers in diameter,
involving 100's of interfaces, making capsid assembly a multiscale process.
While several aspects of icosahedral capsid self-assembly have been
studied in detail \cite{elrad2010encapsulation, reguera2019kinetics}, its
multiscale aspect still remains poorly understood.

\begin{figure}[htpb] \centering
\begin{subfigure}{0.45\textwidth}
\includegraphics[width=\textwidth]{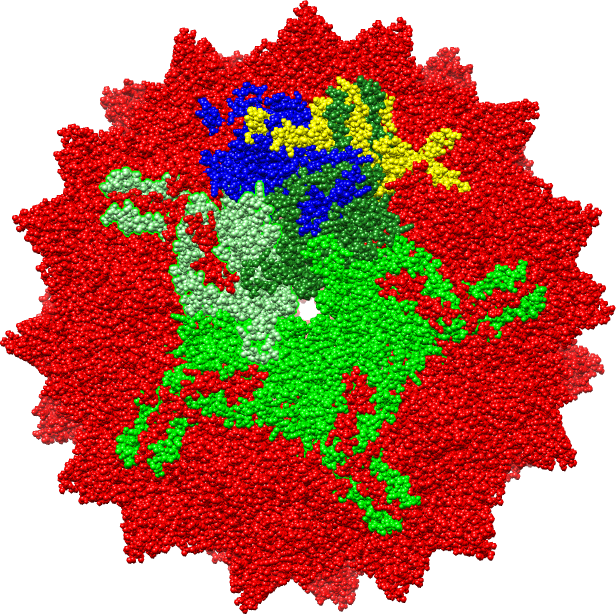}
\caption{\centering}
\label{fig:AAV} 
\end{subfigure}
\begin{subfigure}{0.45\textwidth}
\includegraphics[width=\textwidth]{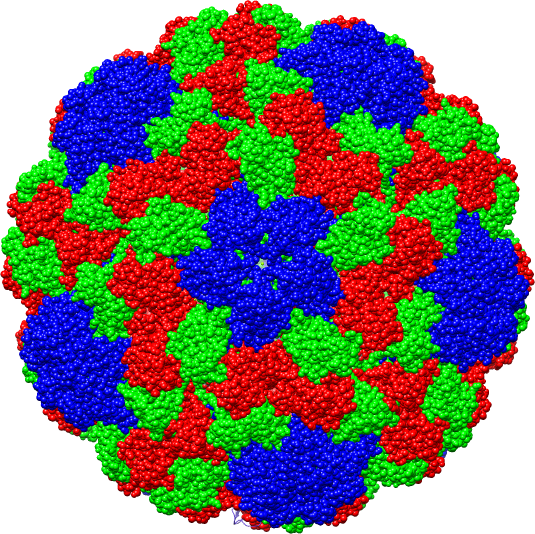}
\caption{\centering}
\label{fig:BMV}
\end{subfigure}
\begin{subfigure}{0.75\textwidth}
\includegraphics[width=\textwidth]{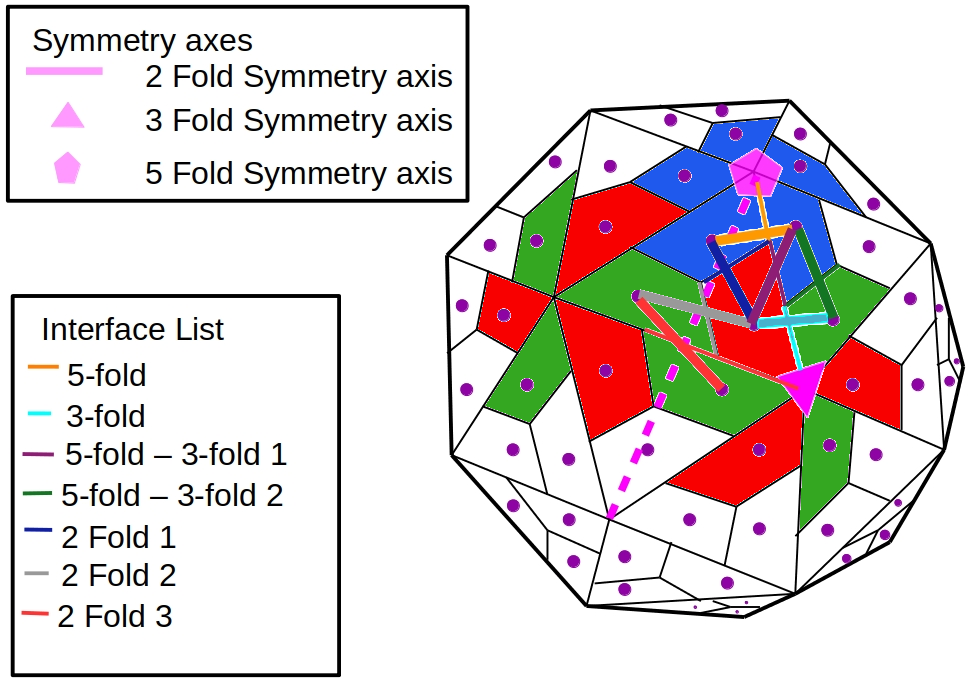}
\caption{\centering}
\label{fig:T=3Interfaces}
\end{subfigure}
\caption{{\bf Structures of a $T=1$ and $T=3$ viruses, and a cartoon
showing the types of VP monomers and interfaces in the former.}
(a) X-ray structure of AAV2 (a $T=1$ virus). All VP monomers are
identical, and the VP monomers colored using the non-dominant colors are used
only to highlight the 3 types of interfaces.  VP monomers at the 5-fold
interface are colored shades of green, light green and blue form a 2-fold
interface assembly, and dark green, blue and yellow pairwise form 3-fold
interface assemblies.  (b) X-ray structure of BMV (a $T=3$ virus) showing 3
types of VP monomers (green, blue, and red).  (c) A cartoon of a $T=3$ virus
showing the 3 types of VP monomers (green, blue, and red), 7 types of
interfaces, and 3 symmetries (shown in pink).  See Section
\ref{sec:introduction}.}
\label{Fig 1}
\end{figure}

Like most other supramolecular assemblies that occur widely in nature, viral
capsid self-assembly is extremely robust, rapid, and spontaneous. Spontaneity
makes it difficult to control in vitro, rapidity makes it difficult to get
snapshots of the process, and robustness makes it difficult to isolate crucial
combinations of assembly-driving inter-atomic interactions (see
\figref{Fig 1}).

Assembly involves two types of interactions: (i) the viral coat protein (VP)
interactions and (ii) the VP-genome interactions. Although the genome or other
biomolecules could influence the VP-VP interaction during the first step of
capsid assembly, understanding VP-VP interactions and the VP intermediates
generated, in system not requiring additional input, can inform the utilization
of viruses for beneficial applications or the generation of assembly inhibitors
that disrupt the formation of pathogenic viruses. Consequently, the
self-assembly of several types of icosahedral, non-enveloped viral capsids from
identical VP monomers is to date an area of major interest. 

A key component in understanding the virus assembly process is identifying
those \emph{crucial hotspot interactions} whose removal disrupts assembly.
Experimental approaches used to measure the forces involved in determining or
orchestrating the VP-VP interaction of the assembled virus include
cryo-electron microscopy and image reconstruction, X-ray crystallography, and a
variety of quantitative interaction proteomic methods
\cite{tuncbag2016potential} which provide high resolution information about the
purified capsid in the crystalline and aqueous states respectively
\cite{luque2020cryo}, complete list found on VIPERdb
(\url{http://viperdb.scripps.edu}). These high-resolution structures can be
used to select residues that are conserved within the virus genus or family and
located within symmetry-related interfaces of the icosahedron.
\emph{Site-directed mutagenesis} of the VP followed by gel filtration, light
scattering, or sedimentation coefficient to measure the size of the VP oligomer
and to determine the effect of the mutagenesis on capsid assembly
\cite{casini2004vitro}. Other methods used to verify capsid assembly include
native capsid immunoblot or enzyme-linked immunosorbent assay (ELISA), and
sometimes cryo-electron microscopy and other techniques for measuring sizes and
concentrations of subassemblies. These methods of capsid assembly prediction
and verification are time consuming and expensive. Additionally, the
predictions may not yield mutants that are critical to the process. Thus, there
is a need for rapid and reliable mathematical and computational tools for
modeling supramolecular assembly that can inform further experimentation,
including resource intensive in-silico experimentation using computational
alanine scanning (CAS) or fine-grained molecular dynamics (MD) that have to be
scaled up from the protein-protein interface level to the capsid level
consisting of at least 150 interfaces.

\medskip\noindent{\sl Contribution.}
The strong influence of entropy contributes to the poorly understood
statistical mechanics of the capsid assembly process, whose free-energy
landscape arises from the system of inter-atomic interactions at interfaces
between the nearly identical VP monomers. Icosahedral symmetry restricts the
interface types to a small set. 

To predict the importance (\emph{cruciality}) of a specific inter-atomic
interaction at an interface for successful capsid assembly, we analyze the
viral capsid assembly landscape at two scales, the \emph{interface-scale} and
the \emph{capsid-scale}.
At the interface-scale, we measure the cruciality of an interaction for
successful subassembly across an interface type by approximating the changes in
the partition function when the interaction is removed (discussed in Section
\ref{sec:backgroundEntropy}). We use two \emph{measures} of change in the
partition function. The first uses the partition function for all the minimal
energy regions, representing all the stable subassembly configurations. The
second uses the normalized partition function for the potential energy basin
corresponding to the specific subassembly configuration occurring in the
successfully assembled capsid. This estimates the probability that a stable
configuration is in fact the successful subassembly configuration. We use the
ratio of each of these quantities with and without an interaction - averaged
over a principled selection of small subassemblies across an interface type -
to measure cruciality of that interaction for that interface type (discussed in
Section \ref{sec:results:ConfPrediction}).

Both measures of change in partition function are rapidly approximated as a
\emph{bar-code} that abbreviates the \emph{atlas} of the interface assembly
landscape. The atlas is generated - with minimal sampling - by the geometric
method and curated opensource software EASAL (efficient atlasing and search of
assembly landscapes \cite{Ozkan2011, Ozkan:toms, maineasal}). The input to
EASAL consists of (a) the VP monomer geometry - atom coordinates; and for each
interface type, (b) pair-potentials for a candidate set of assembly-driving
interactions along with Van der Waals sterics, and (c) small subassembly
structures extracted from known capsid structures. An \emph{atlas} is a
partition of the assembly landscape into contiguous region of nearly
equipotential energy called \emph{active constraint regions} or
\emph{macrostates} (discussed in Section \ref{sec:backgroundEASAL}), organized
as a refinable, queryable roadmap, that can further be abbreviated as a
\emph{bar-code}. The constraints are the pair-potentials as in (b) above. The
active constraint graphs are analyzed using combinatorial graph rigidity,
whereby the effective dimension of a macrostate becomes a proxy for its energy
level. The methodology gives fast, light-weight algorithms (100 to 1000 times
faster than prevailing methods \cite{maineasal, easalSoftware, ozkan2014fast})
with rigorously proven accuracy-efficiency tradeoffs.

We additionally give two \emph{types} of predictions, each validated by
in-vitro, site-directed mutagenesis results found in literature
\cite{wu2000mutational, okinaka2001c, bleker2005mutational, reguera2004role,
lochrie2006mutations, riolobos2006nuclear, UFDC000005707}. We note that the
biophysical assays for AAV2 were carried out by the Mavis Agbandje-Mckenna's
lab \cite{UFDC000005707} contemporaneously with the development of our
computational model and prediction. Our first direct, ab-initio prediction
generalizes an interface-scale prediction to the capsid-scale by assuming equal
importance of each type of interface for capsid assembly and without explicitly
accounting for kinetics. {\sl Despite this assumption, and despite the
prediction being completely blind to the in-vitro site-directed mutagenesis
data} used for validation, this direct, interface-scale prediction
correlated well with site-directed mutagenesis data towards capsid assembly
disruption in 3 viruses (see the figures in Section \ref{sec:twoScale1}).

Our second prediction additionally incorporates a capsid-scale prediction of
the cruciality of an interface for capsid assembly. The dimension of the capsid
assembly landscape - involving several VP monomers (60 for $T=1$ and 180 for
$T=3$) - makes direct computations intractable. Hence, we treat the capsid as
being recursively assembled from stable subassemblies at interfaces
\cite{sitharam2006modeling, sitharam:Assembly}, where subassemblies are
intermediate oligomeric structures of the capsid, assembled at interfaces
(subassemblies are formally defined in Section \ref{sec:combinatorialEntropy}).
The likelihood of such an \emph{assembly tree}, given successful capsid
assembly, is a measure of combinatorial entropy (discussed in Section
\ref{sec:combinatorialEntropy}). This depends both on the stability and
formation rates of the intermediate subassemblies, and the number of equivalent
assembly trees under icosahedral symmetry \cite{sitharam2005counting,
bona2008influence, bona2011enumeration}. The cruciality of an interface for
successful capsid assembly is then determined by all the assembly trees that
involve that interface. The relative weights of the cruciality measures
described above - the bar-code measuring change of partition function at the
interface-scale, and the combinatorial entropy at the capsid-scale - are then
determined by statistical learning (discussed in Section \ref{sec:microscale}).
The learning algorithm uses - for training - a small fraction of the
mutagenesis and biophysical assay data towards assembly disruption, to learn
the parameters of the statistical model. The remainder of the mutagenesis data
is used to validate the cruciality of residues for capsid assembly of 3 types
of viral capsids, Adeno Associated Virus serotype 2 (AAV2), which is
non-pathogenic, and Minute Virus of Mice (MVM), which is pathogenic to mice,
both $T=1$ single stranded DNA viruses, and Bromo Mosaic Virus (BMV), a $T=3$
single stranded RNA virus, pathogenic to both monocotyledon and dicotyledon
plants. 

\begin{figure}[htpb]
\centering
\includegraphics[width=\textwidth]{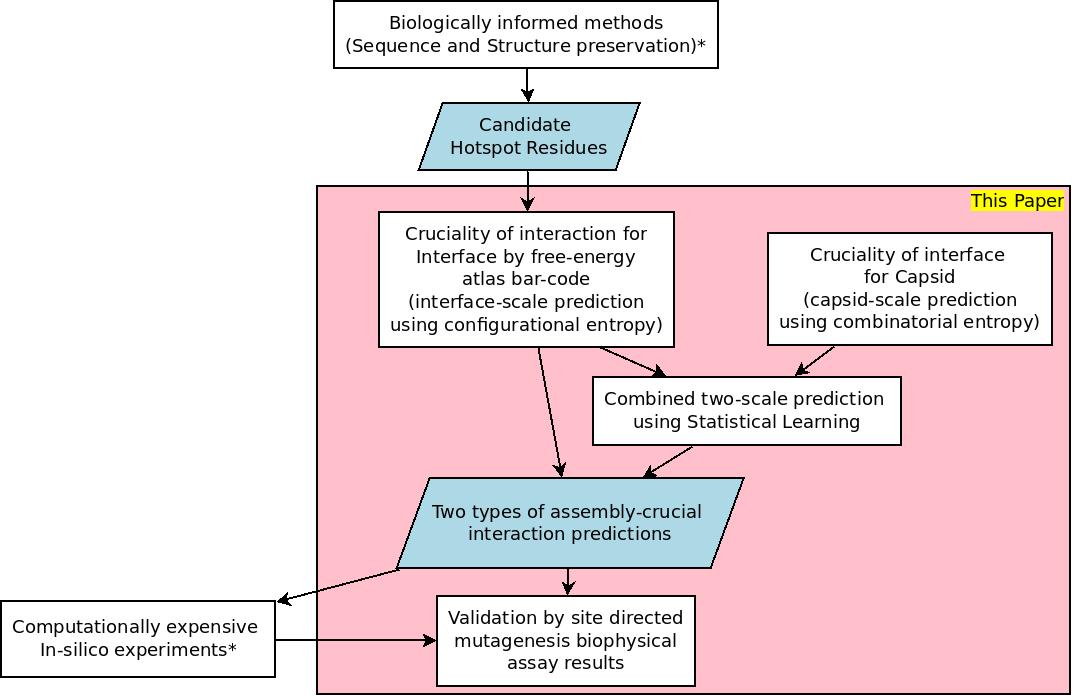}
\caption{\bf{Flow chart of the methodology in this paper and connections to
existing methods.} See Section \ref{sec:introduction} and Section \ref{sec:Motivation}.}
\label{Fig 2}
\end{figure}

Our predictions are reproducible using our curated opensource software EASAL
(efficient atlasing and search of assembly landscapes
\url{http://bitbucket.org/geoplexity/easal}, see also video
\url{https://cise.ufl.edu/\~sitharam/EASALvideo.mpeg}, and user guide
\url{https://bitbucket.org/geoplexity/easal/src/master/CompleteUserGuide.pdf}).
Figure \ref{Fig 2} summarizes our overall approach.

Overall, the emphasis of this paper is not the comparison of our
interface-scale or capsid-scale predictions with prevailing methods for each
individual scale.  Rather, our emphasis is on the novel conceptual
underpinnings of each of our single-scale predictions, and the  validation,
using in-vitro site directed mutagenesis data, of our synthesized two-scale
predictions. As Figure \ref{Fig 2} shows, different aspects of our
method can be mixed and matched with prevailing methods to leverage
complementary strengths (e.g., for interface-scale hotspot prediction, or
extensive in-silico validation).

We are unaware of any previous study that spans the range from multiscale
statistical mechanical predictions of crucial hotspot inter-atomic interactions
to validation using site-directed mutagenesis results. Previous studies use
coarse-grained single-scale, or knowledge-based predictions or use
resource-intensive in-silico validation via Computational Alanine Scanning
(CAS) or via fine-grained Molecular Dynamics (MD) (see detailed discussion in
Section \ref{sec:Motivation}). By reliably and rapidly (taking no more than 1.5
hours per interface on a laptop with Intel Core i5-2500K @ 3.2 Ghz CPU
and 8GB of RAM) narrowing down target interactions, our predictions can inform
and reduce the time spent on time-consuming in-vitro and in-vivo experiments,
as well as more computationally intensive in-silico analyses.

\subsection{Related work}
\label{sec:Motivation}
At the interface-scale there are several types of methods for predicting
crucial hotspot protein-protein interactions (PPI). All of these methods (like
ours) use as input a shortlist of candidate hotspot interactions selected using
evolutionary sequence or structure preservation. Many of the methods are based
on computational alanine scanning (CAS) surveyed recently in
\cite{Ibarra2019hotspot}. CAS in turn uses Monte Carlo (MC) or Molecular
Dynamics (MD) simulations \cite{rajamani2004anchor,
xia2010apis,steinbrecher2017predicting} to compute binding affinity, free
energy or entropic influences of the hotspots. Other methods additionally use a
combination of shape specificity, solvent accessibility and prior
knowledge-base of PPI via different types of statistical inference or machine
learning, see e.g., \cite{darnell2007decision, zhu2011kfc2, wang2012prediction,
wang2014prediction,ye2014prediction, deng2014predhs, sukhwal2015ppcheck,
sun2016accurate, hu2017protein, murakami2017network, Wang2018,
barradas2018structural, liu2018machine}. While exhaustive CAS methods are
sometimes validated by site-directed mutagenesis data, most of the hotspot
predictions are validated by in-silico CAS experiments, e.g., the SKEMPI
database \cite{jankauskaite2019skempi}. For example, in a recent paper
\cite{Diaz-Valle2019hotspot}, interface-scale hotspot predictions, combined
with specific sequence and structure conservation, have been directly
extrapolated to viral capsid scale hotspot predictions. Validation, however,
was through computationally intensive all-atom MD sampling. In Figure
\ref{Fig 2}, this approach involves the boxes marked by *.

Free-energy landscapes of protein-protein interface assembly are driven by weak
inter-atomic forces and non-covalent bonds and are strongly influenced by the
configurational entropy. However, full-blown computation of configurational
entropy is a notoriously difficult problem. All prevalent methods for
configurational entropy computation rely on computing the volumes of assembly
landscape regions, typically by MC or MD sampling \cite{kaku, head1997mining,
rapaport1999supramolecular, reddy1998energetics, andricioaei2001calculation,
HAGAN200642, gfeller2007uncovering, hnizdo2007nearest, killian2007extraction,
hnizdo2008efficient, Zhou_Gilson_2009, king2012efficient}, which is prohibitive
due to the high dimension of the assembly landscape. High geometric or
topological complexity of the assembly landscape (disconnectedness, channels of
varying effective dimension etc.), means that sampling techniques like MC or MD
can only claim stochasticity and uniform sampling in the limit, i.e., when they
run for sufficiently long or start from sufficiently many initial
configurations \cite{hensen2010estimating, fogolari2015distance, huang2011free,
dunton2014free, staneva2011binding, prada2009exploring}. Some works such as
\cite{varadhan2006topology,gfeller2007uncovering,lai2009uncovering} infer the
topology of the configuration space starting from MC and MD trajectories and
use topology to guide dimension reduction. On the other hand, methods based
on principal component analyses of the co-variance matrices from a trajectory
of samples in internal coordinates generally overestimate the volumes of
assembly landscape regions. For these reasons, formal accuracy-efficiency
tradeoffs are not provided. In contrast, EASAL uses the novel geometric idea of
convexifying \emph{Cayley} parameters to represent macrostates, and avoids
gradient descent used by the above-mentioned methods, thereby significantly
reducing discarded samples and increasing efficiency. Moreover, the EASAL
method is able to approximate the configurational entropy of small assembly
systems using an atlas bar-code, without relying heavily on sampling, thereby
ameliorating the curse of dimension, as shown using rigorous complexity
analysis and computational experiments \cite{maineasal}; furthermore formal
accuracy-efficiency tradeoff guarantees are provided.

Ab initio methods such as \cite{Chirikjian201199}, based on geometric algebras
are used to give bounds or approximate configurational entropy without
relying on Monte Carlo or Molecular Dynamics sampling. However, it is not clear
how to extend them beyond restricted assembly systems such as a chain or loop
of rigid molecular components, each component consisting of at most 3 atoms,
non-covalently bound to their neighboring components at exactly 2
sites. EASAL on the other hand is applicable more generally to assemblies with
larger inputs.

While for small assemblies it is possible to atlas the assembly landscape and
compute the entropy directly, for larger assemblies, such as virus capsid
systems (consisting 60 VP monomers for $T=1$ and 180 VP monomers for $T=3$
viruses), the assembly landscape is too big to be atlased directly, although 
all-atom MD simulations of viral capsid life-cycle processes (docking etc.)
post-assembly in the literature, e.g., \cite{Perilla2016, Durrant2020}. Therefore,
to tractably deal with the high dimension of their assembly landscape, larger
assemblies are typically treated as being recursively assembled as an interface
assembly system, from a small number of stable intermediate subassemblies
\cite{sitharam2006modeling}.

Several statistical mechanical approaches, as surveyed in
\cite{statisticalMechanics2015, virusMechansim2015}, could be said to combine
configurational entropy and combinatorial entropy into a single scale to
analyze kinetics \cite{zlotnick1994build, zlotnick1999theoretical,
zlotnick2000mechanism, endres2002model, ZLOTNICK2003536,
zlotnick2005theoretical, Hagan2010, Bajaj-Quantified, schlicksup2018hepatitis}.
The assembly model \cite{schwartz} based on the local-rules theory
\cite{berger1994local, berger1994mathematics, berger1995local,
schwartz1998local} computes the combinatorial entropy considering both the
number of different assembly pathways and the kinetics at each assembly stage.
However, such single-scale models rely crucially on the simplified
representation of the VP monomers and their geometric interactions, and feature
kinetics, rates and concentrations of subassemblies prominently in their
analyses. The assembly model in \cite{reguera2019kinetics} analyzes the
efficiency and kinetics of the capsid assembly process. However, they use a
simplified coarse-grained model, which assumes that identical capsomeres
(pentamers and hexamers) assemble together to form the entire capsid.
Similarly, the assembly models in \cite{rapaport2004self, rapaport2008role,
elrad2010encapsulation} use truncated capsomeres as sub-units of assembly,
which when assembled give a perfect icosahedron.

While our method does not {\sl separately} model kinetics, it combines
interface-scale and capsid-scale analyses of free-energy, configurational and
combinatorial entropy, which could affect kinetics. Furthermore, our method
does not rely on a single, capsid-scale analysis with simplified
representations of VP monomers, but uses a multiscale analysis.

Several computational studies have been conducted on various aspects of the
virus life cycle. The paper \cite{polles2013mechanical} uses rigidity analysis
on the fully assembled capsids of icosahedral proteins to identify functional
units of the capsid. Assembly pathways have been used to study the
self-assembly of polyhedral systems from identical sub-units; for example, the
paper \cite{pandey2014self} studies the role of assembly pathways and the degrees
of freedom of intermediate subassemblies in the self-assembly of polyhedra with
known isomers. The goal is to manipulate the degrees of freedom of the
intermediate sub-assemblies to increase the concentration of one isomer over
others. In contrast, we use graph and symmetry analysis of assembly pathways of
viral capsids with the goal of identifying interfaces that are crucial to
assembly. We further use graph rigidity to analyze and synthesize the two
scales, namely configurational and combinatorial entropy, of capsid assembly.

\medskip\noindent \textbf{Organization}: The paper is organized as follows.
Section \ref{sec:methods} describes the predictions of cruciality of
inter-monomeric interactions for interface assembly and for capsid assembly as
a whole. Section \ref{sec:results} provides the results validating the
cruciality prediction of interactions to the capsid, in three viruses, AAV2,
MVM, and BMV.

\section{Materials and methods}
\label{sec:methods} 
In Section \ref{sec:backgroundEntropy}, we provide some background on the
configurational entropy of virus capsid assembly. In Section
\ref{sec:backgroundEASAL}, we describe key features of the EASAL methodology (
see software \url{http://bitbucket.org/geoplexity/easal}, video
\url{https://cise.ufl.edu/\~sitharam/EASALvideo.mpeg}, and user guide
\url{https://bitbucket.org/geoplexity/easal/src/master/CompleteUserGuide.pdf}). In
Section \ref{sec:combinatorialEntropy} we discuss the combinatorial entropy in
viruses.

In Section \ref{sec:nanoscale} we describe the computation of the cruciality of
inter-atomic interactions across VP monomers to interface subassembly and
thereby to capsid assembly. In Section \ref{sec:microscale} we describe a
second scale of cruciality of interfaces to capsid assembly. In Section
\ref{sec:multiscale} we describe statistical models to combine the
interface-scale configurational entropy and the capsid-scale combinatorial
entropy to predict the cruciality of an interaction at the capsid level.

\subsection{Background: configurational entropy in virus assembly}
\label{sec:backgroundEntropy}
The efficacy of viral capsid assembly is largely due to the structure of its
equilibrium free energy landscape. Specifically, the depth and volume of the
potential energy basins, including the basin containing the successfully
assembled capsid configuration. The free energy at a basin depends on the
average potential energy and the configurational entropy of the basin. Of
these, the computation of the configurational entropy dominates the computation
of free energy.

Let $E(x)$ be the potential energy function, defined over the assembly
landscape, for an assembly configuration $x$ (the function $E$ is described in
detail in Section \ref{sec:backgroundEASAL}). The partition function $Q$ is a integral
over the energy basin $\beta$, given by $$ Q = \int_\beta e^\frac{-E(x)}{k_B~T}
\text{d}x $$ where $x \in \beta$ is a configuration in the basin, $k_B$ is the
Boltzmann's constant and $T$ is the absolute temperature. The configurational
entropy $S$ of the basin is $$ S = k_B \ln Q + \frac{\langle E \rangle}{T} $$
where $\langle E \rangle$ is the the average energy over the basin.

The free energy $F$ of a system with a single energy basin $\beta$ is given by:
$$ F = \langle E \rangle -TS$$

Hence, over a region $C$ of constant energy $E_C$, for example an active
constraint region as defined in EASAL, the entropy is merely a function of the
volume $V_C$ of the region.
\begin{equation}
\label{eq:entropy_c}
S_C = k_B \ln V_C
\end{equation}
where $V_C = \int_C \text{d}x$.


In a landscape with multiple potential energy basins $\beta_i$, each of which
has a constant energy $E_i$, the partition function of each energy basin $Q_i$
can be expressed as a weighted sum of the volumes $V_i$ of the different
basins.

\begin{equation}
Q_i = \int_{\beta_i} \text{d}x \cdot e^\frac{-E_i}{k_B T} = V_i \cdot
e^\frac{-E_i}{k_B T}
\label{eq:partition_function}
\end{equation}

The normalized partition function $p_i$ is the probability of finding
the system in the energy basin $\beta_i$:
\begin{equation}
p_i = \frac{Q_i}{\sum_i Q_i}
\label{eq:normalized_partition_function}
\end{equation}
In the next section we show how to approximate the computation of the partition
function by generating an atlas of the capsid assembly landscape using EASAL and extracting
a relevant bar-code.
\subsection{Atlasing and entropy computation using EASAL}
\label{sec:backgroundEASAL}
An \emph{interface assembly system} (see \figref{Fig 1}) consists of
(a) the VP monomer geometry - atom coordinates; and for each interface type, (b)
short-range Lennard-Jones potentials for a candidate set of interactions, i.e.,
atom pairs (one from each VP monomer) along with Van der Waals sterics, and (c)
small subassembly structures extracted from successfully assembled capsid.

The potential energy $E(x)$ for an \emph{interface assembly configuration} $x$
has one Lennard-Jones term for each atom pair (one from each VP monomer). In
EASAL, the short-range Lennard-Jones pair potentials are \emph{geometrized} by
discretizing into three intervals: large distances at which Lennard-Jones
potentials are no longer relevant, contributing $E_h$ to the potential energy
of the configuration; short distances prohibited by Van der Waals forces; and
interval between the two known as the Lennard-Jones \emph{well}, contributing
$E_l$ to the potential energy of the configuration. We say that a pair of atoms
has an \emph{active constraint} if the distance between their centers is within
the discretized Lennard-Jones well.

For a landscape with $N$ Lennard-Jones terms, potential energy of a
configuration with $N_a$ active constraints is given by:

\begin{equation}
E = N_a E_l + (N-N_a) E_h = N E_h - N_a(E_h - E_l)
		\label{eq:EnergyDiff}
\end{equation}

In the expression for partition function in Eq. \ref{eq:partition_function},
each configuration contributes a weight

\[
e^\frac{-E}{kT} = e^\frac{-N E_h + N_a(E_h - E_l)}{kT} = C\cdot (e^\frac{E_h
- E_l}{kT})^{N_a}
\]
where $C = e^\frac{-N E_h}{kT}$ is a constant of the landscape and is
canceled out when calculating the normalized partition function,
and the weight

\begin{equation}
w(N_a) = (e^\frac{E_h - E_l}{kT})^{N_a}
		\label{eq:weights}
\end{equation}

\begin{figure}[htpb]
\centering
\includegraphics[width=\textwidth]{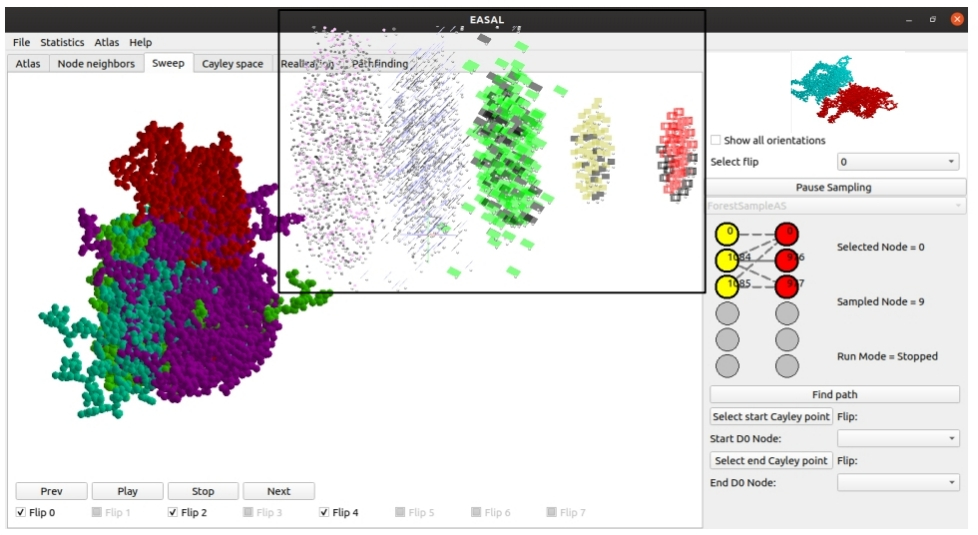}
\caption{{\bf Screenshot of the EASAL software showing all configurations in an
active constraint region in the atlas of the interface assembly system of the
two VP monomers shown on top right.}  The region's active constraint graph is
shown at bottom right, with red and yellow representing atoms in different VP
monomers, the single bold edge representing a single active constraint or
interaction $c$, and the dashed lines representing the 5 Cayley parameters that
are used to convexify this effectively 5-dimensional region. On the main
screen, the red VP monomer is held fixed and all of the second VP monomer's
relative positions (satisfying the one active constraint $c$) are shown. The 3
different colors (cyan, green and purple) of the second VP monomer sweeps
represent distinct orientations within the same active constraint region.
(inset) Atlas with changes when an interaction is disabled. Active constraint
regions (nodes of the atlas) of different dimensions are shown in different
colors, with red nodes representing regions with 2 active constraints, or 4
effective dimensions, and each of the successive strata (from right to left)
showing regions of one more active constraint, or one lower energy level or
effective dimension. The left most are the 0-dimensional or lowest energy
regions, each of which is the bottom of a potential energy basin, with all its
ancestor regions participating in the basin.  The black nodes are the active
constraint regions that disappear from the atlas due to the removal of a
candidate inter-atomic interaction.  See Section \ref{sec:backgroundEASAL} and
Section \ref{sec:nanoscale}.} 
\label{Fig 3}
\end{figure}

With this geometrization of energy, the potential energy basin is completely
determined by its partition into {\em active constraint regions}, i.e., regions
of the assembly landscape whose configurations have a particular set of active
constraints and hence, nearly constant potential energy. This gives a queryable
\emph{roadmap} of the basin, where each region is uniquely labeled by an
\emph{active constraint graph}, whose edges are the active constraints and
whose vertices are the participating atoms (see \figref{Fig 3}).
Using combinatorial rigidity \cite{SJS:Handbook}, each active constraint
generically reduces the effective dimension of the region by one. The bottom of
each basin is a 0-dimensional region $R$, with active constraint graph $G$,
containing the minimum energy configurations. The higher energy regions leading
to $R$ are exactly those that have active constraint graphs that are subgraphs
of $G$.

One of EASAL's key features is the generation of roadmaps for all basins,
called an \emph{atlas}, without relying heavily on sampling. This is achieved
by using a recursive method that searches the interior of higher energy regions
for boundary regions with exactly one new active constraint. Searching for such
boundary regions (which are effectively of one fewer dimension) has a higher
chance of success than directly looking for the lowest energy regions, which
are the lowest dimensional active constraint regions.

Staying within active constraint regions is achieved by a second key feature of
EASAL: \emph{convexifying} active constraint regions using customized,
distance-based or \emph{Cayley} parametrization, avoids gradient-descent to
enforce active constraints, and results in high efficiency search with minimal
sampling and reduced repeated or discarded samples. In addition, it is
straightforward to compute the inverse map from the Cayley parameter values to
their corresponding finitely many Cartesian configurations. Altogether, EASAL
obtains comparable coverage with 100 to 1000 times fewer samples than
prevailing methods \cite{maineasal,easalSoftware,ozkan2014fast}. Cayley convexification
leverages geometric features that are unique to assembly (as opposed to protein
folding). Together, the active constraint regions, their effective dimensions,
and their volume approximations obtained through Cayley parameterization,
provide an abbreviated atlas \emph{bar-code} for the basin structure of the
assembly landscape.

\subsection{Background: combinatorial entropy in virus assembly}
\label{sec:combinatorialEntropy}
Combinatorial entropy of capsid assembly captures the number of possible
ways in which a successful assembly configuration can be recursively decomposed
into subassemblies down to the rigid motifs in the VP monomers
\cite{sitharam2006modeling, bona2011enumeration, bona2008influence,
sitharam2005counting}. In reverse, larger assemblies are treated as being
recursively assembled as interface assembly systems. Since the VP monomers that
are far away from the interface tend to have little impact on the assembly, we
can simplify the participants of each interface assembly system to VP monomers or
dimers near the interface.

\begin{figure}[htpb]
\centering
\includegraphics[width=\textwidth]{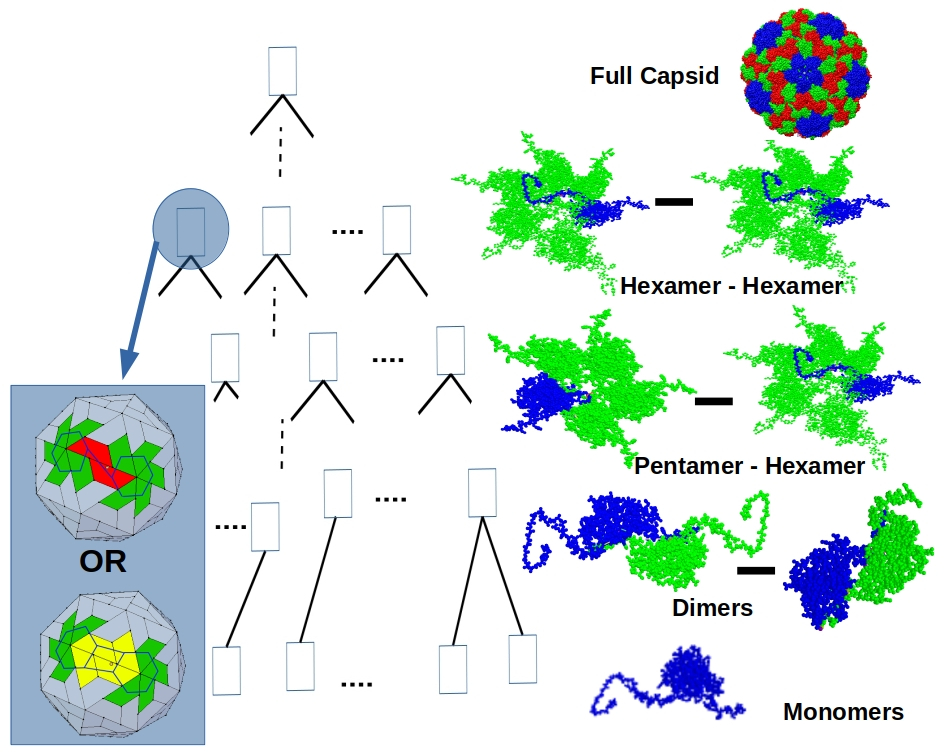}
\caption{ {\bf An assembly tree of a $T=3$ viral capsid.} The root node represents a
successfully assembled viral capsid. Each internal node represents an interface
assembly system that contains a stable subassembly configuration that is part
of the known, successfully assembled capsid configuration. Children of a node
are the participating multimers for the node’s interface assembly system. The
leaf nodes represent the VP monomers. To the right of the nodes are their
candidate stable subassembly configurations taken from the $T=3$ BMV X-ray
capsid structure. At internal nodes, a choice is made between multiple
candidate interface assembly systems. On the left we highlight an internal node
with 2 available choices for hexamer-hexamer interfaces, of which one is
chosen: the inset shows the choices - a single VP dimer interface highlighted
in red; and two VP dimer interfaces, highlighted in yellow. See Section
\ref{sec:combinatorialEntropy}. }
\label{Fig 4}
\end{figure}

As shown in \figref{Fig 4}, there is typically more than one way of
treating a subassembly as an interface assembly. When there are multiple
interfaces to choose from, we consider the free energy and reaction rates of
each of the options and pick the best interface for the subassembly.

With this setup, we define a labeled binary tree, called an \emph{assembly
tree}, to describe how a series of subassemblies leads to a full capsid
assembly. In an assembly tree, the root node is a successfully assembled viral
capsid, and the leaves are VP monomers. Every internal node of the tree is a
subassembly, labeled by its best interface (as defined earlier).
\figref{Fig 4}, shows an assembly tree for a $T=3$ viral capsid.
Given the free energy and reaction rate of each subassembly and the structure
of the assembly tree, we can define its likelihood under the assumption of
successful assembly.

An \emph{assembly pathway} is a collection of assembly trees that satisfy some
prediction-related criteria \cite{bona2011enumeration, bona2008influence,
sitharam2005counting}. For example, all assembly trees that are in one
equivalence class under icosahedral symmetries can be grouped as a single
assembly pathway. As another example, an assembly pathway can be defined as the
collection of such symmetry classes that do not use specific types of
interfaces. The papers \cite{sitharam2005counting, bona2008influence,
bona2011enumeration} enumerate assembly pathways and compute their likelihood
for such criteria.
\subsection{Interaction cruciality at interface-scale}
\label{sec:nanoscale}
We use the atlas generated by EASAL to compute two quantities for each
interface assembly landscape: (a) the partition function for minimal energy
regions (basin bottoms), and (b) the normalized partition function for the
potential energy basin corresponding to the known (successful) interface
subassembly configuration called the \emph{true realization}. These two
parameters serve as an atlas bar-code to determine the cruciality of
interactions at the interface-scale.

\begin{figure}[htpb]
\centering
\includegraphics[width=0.9\textwidth]{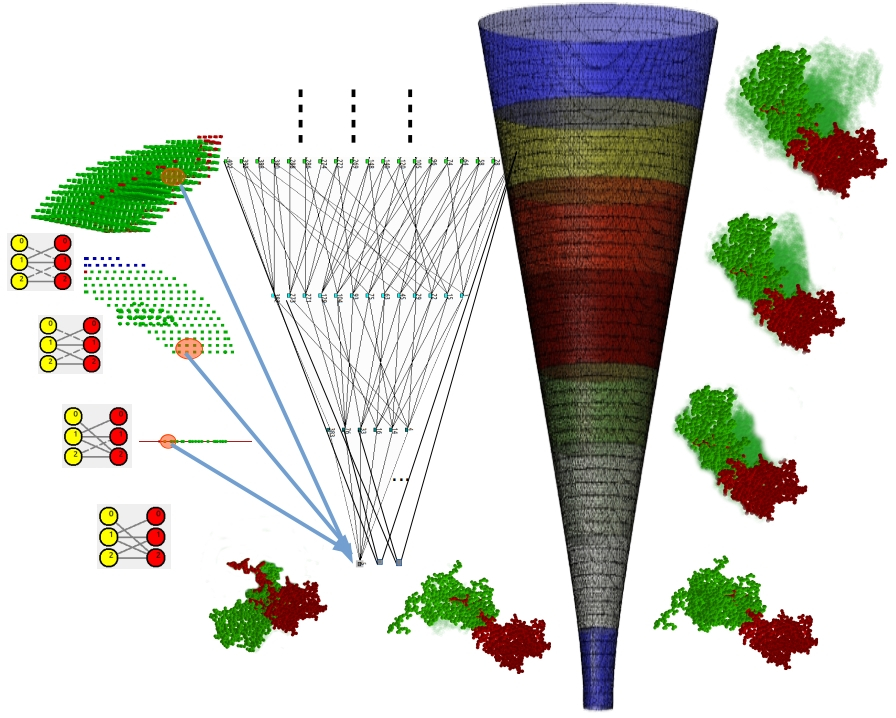}
\caption{ {\bf Prediction using \emph{cruciality bar-codes} of Section
\ref{sec:nanoscale} for two VP monomers assembling across a hexamer interface
in BMV.} Each node in the atlas roadmap in the middle represents an active
constraint region (macrostate) in EASAL. Example active constraint graphs are
shown at far left: the yellow and red circles represent atoms participating in
active constraints (interactions) in the two VP monomers. At each successive
level, the number of active constraints increases by 1 and the energy level and
effective dimension decrease by 1. The atlas nodes in the bottom-most row
represent the 0-dimensional, lowest energy, stable assembly configurations;
example configurations shown below them. Their total number (for a given
interface $s$, on removal of a given interaction or constraint $r$) gives
$\nu^{r,s}_{minima}$ in the computation of the cruciality bar-code. Each such
configuration together with nearby higher-energy configurations in all of their
ancestor nodes constitute one potential energy basin. Their sum, across all
basins, weighted by energy level gives the denominator of $\nu_{capsid}^{r,s}$.
The rightmost of the stable assembly configurations at the bottom corresponds
to the \emph{true realization}. Above it, the 3 solid configurations and the
transparent sweeps around them show the closest configurations to the true
realization in successively higher energy regions in its basin (one region each
for 3 energy levels shown). To the far left, these sweeps are shown as orange
highlights in the corresponding Cayley parameterized regions. The colorful
basin plot shows the total weighted configurations in the true basin,
stratified by dimension or energy level. Their sum is the numerator of
$\nu_{capsid}^{r,s}$.}
\label{Fig 5}
\end{figure}

As mentioned earlier, the bottom of each basin is a 0-dimensional region $R$,
with active constraint graph $G$, containing the minimum energy configurations.
The higher energy regions leading to $R$ are exactly those that have active
constraint graphs that are subgraphs of $G$. \figref{Fig 5}
illustrates, using EASAL screenshots, the basin structure of two
VP monomers assembling across a hexamer interface in BMV.

Two assembly configurations are considered distinct if and only if their
\emph{similarity} distance (the 2-norm distance between their point coordinate
vectors) is at least $\varepsilon$. The number of distinct Cartesian
configurations in a region then becomes an approximate measure of the size or
volume of the region (configurational entropy associated with that region).

For any interface assembly system $s$, since the energy of all 0-dimensional
configurations is the same, we approximate the sum of the $Q_i$'s in
Eq. \eqref{eq:partition_function}, with the number of distinct configurations in
the union - denoted by $R^s_{0}$ - of all the 0-dimensional active constraint
regions. Formally, the \emph{partition function for all minimal energy
regions} of the atlas of a given interface assembly system $s$ is denote it by
\begin{equation}
\nu_{minima}^s := |R_0^s|
\label{eq:nu_minima}
\end{equation}
The approximation to the normalized partition function of Eq. \eqref{eq:weights} is
the ratio of the number of distinct 0-dimensional configurations in the basin
of the true configuration (we call this set $R^s_{true}$) to $\nu_{minima}^s$.
This approximates the probability that the assembly process ends in the true
configuration. To improve this approximation we weight each configuration $x$
inversely to its proximity to a higher energy region $A$ by the weight
$w(N_a(x))$ of Eq. \eqref{eq:weights}, where $N_a(x)$ is now the number of active
constraints in the configurations in $A$.

Thus the normalized partition function for the potential energy basin
corresponding to a successful interface assembly configuration is computed
using Eq. \eqref{eq:normalized_partition_function} as follows:
\begin{equation}
\nu_{capsid}^s := \frac{\sum_{\x \in {R^s_{true}}} w(N_a(\x))}{\sum_{\x \in
{R^s_{0}}} w(N_a(\x))}
		\label{eq:nu_capsid}
\end{equation}
Finally, these quantities are used to define our measure of \emph{cruciality of
a given input inter-atomic interaction $r$ for a given interface assembly
system $s$} to result in a given true configuration. First we define
$\nu_{minima}^{r,s}$ and $\nu_{capsid }^{r,s}$ as the same quantities in
Eq. \eqref{eq:nu_minima}, and Eq.
\eqref{eq:nu_capsid}, respectively, obtained by restricting to a portion of the
atlas, i.e., those regions where $r$ is not an edge in the active constraint
graph (see \figref{Fig 3} and \figref{Fig 5}). Now, the
\emph{cruciality bar-code} is defined as: 
\begin{equation} (\mu^{r,s}_{minima}
:= \frac{\nu_{minima}^{r,s}}{\nu^s_{minima}}, \mu^{r,s}_{capsid} :=
\frac{\nu_{capsid}^{r,s}}{\nu^s_{capsid}}) 
\end{equation}

\specialheads{Accounting for multimers assembling at an interface}
In a capsid assembly tree, the subassembly at an interface could involve either
a VP monomer pair or a multimer pair. Although the pair potentials at the
interface are specified between the VP monomers closest to the interface, each
VP monomer could be part of a multimer whose atoms influence the interface
assembly landscape through Van der Waals sterics. We have found that for
larger multimers the steric contribution from the VP monomers far from the
interface is negligible and that it is sufficient to consider those interface
assembly systems involving certain VP monomer-dimer pairs selected as follows.

\begin{figure}[htpb]
\begin{subfigure}{0.45\textwidth}
\centering
\includegraphics[width=.9\textwidth]{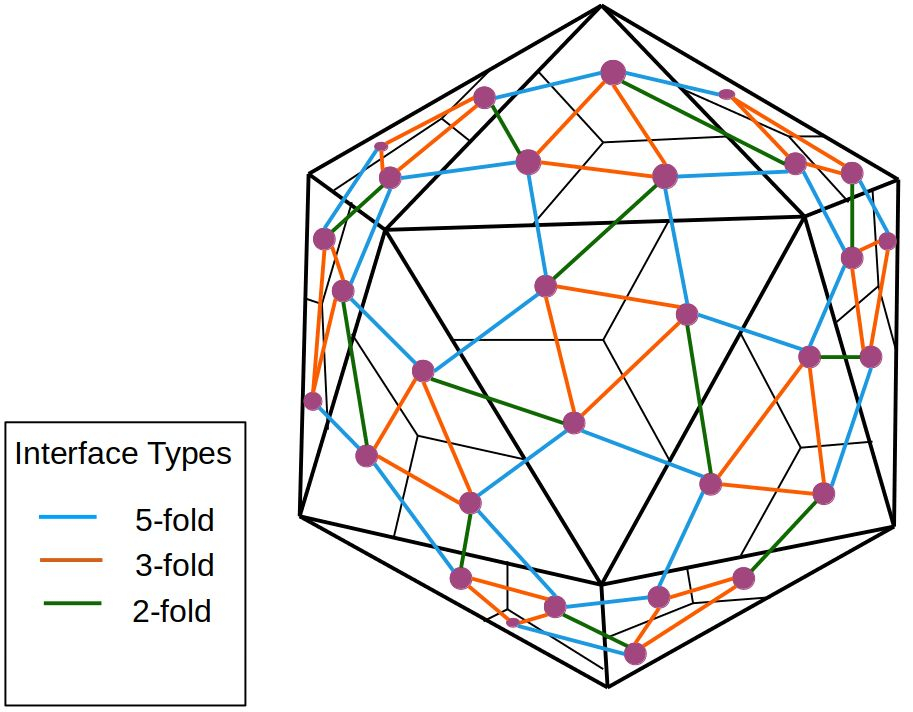}
\caption{\centering}
\label{fig:T1All} 
\end{subfigure}
\begin{subfigure}{0.45\textwidth}
\includegraphics[width=\textwidth]{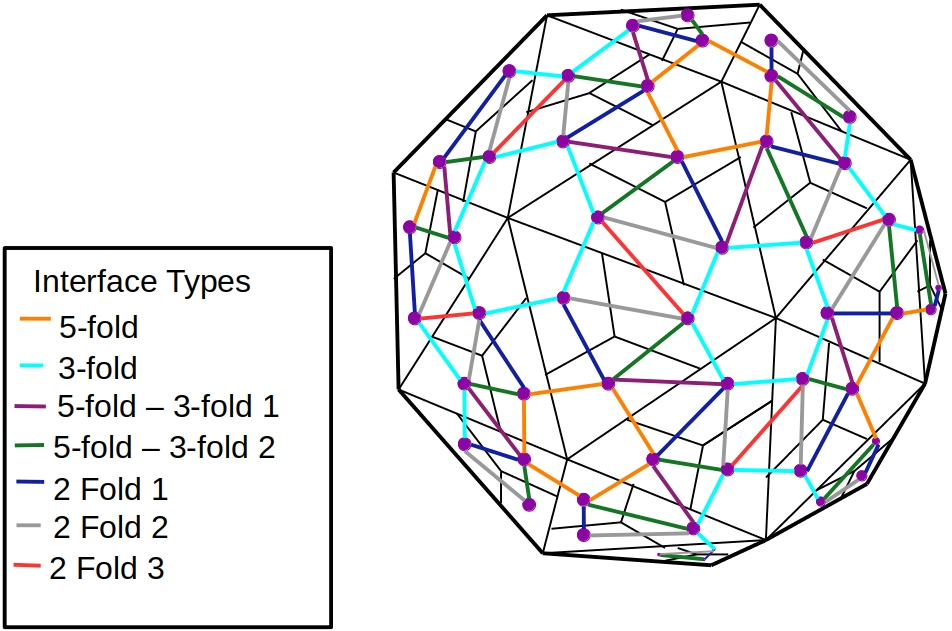}
\caption{\centering}
\label{fig:T=3All} 
\end{subfigure}
\caption{{\bf The dual graphs of $T=1$ and $T=3$ capsid polyhedra.} Faces
of the capsid polyhedra are shown with black edges and the colored edges
give the dual graph. See Section \ref{sec:nanoscale}.
}
\label{Fig 6}
\end{figure}

The \emph{dual graph} of a virus capsid is obtained from the icosahedrally
symmetric \emph{capsid polyhedron}, with one face per VP monomer, where
interfaces are represented by adjacent faces (see
\figref{Fig 6}). There is
one vertex of the dual graph corresponding to each face of the capsid
polyhedron and an edge between two vertices if the corresponding faces 
share an interface.

For the interface represented by the edge $ab$ in the dual graph, we consider
each triangle $abc$, and generate 3 atlases with the following assembly systems
$s$: (i) with VP monomers $a$ and $b$, (ii) VP dimer $ac$ and VP monomer $b$, (iii) with
VP dimer $bc$ and VP monomer $a$. For the three $T=1$ interface types, this gives 9
assembly systems, and for the seven $T=3$ interface types, this gives 31 assembly
systems.

Now, $\mu^{r,s}_{minima}$ and $\mu^{r,s}_{capsid}$ are computed using the
atlases for the 3 assembly systems $s$ for the same interface $ab$, and then
averaged to get cumulative values. These are denoted $\mu^r_{minima}$ and
$\mu^r_{capsid}$ and are used to measure the cruciality of the interaction $r$ to
the interface $ab$.

\subsection{Interface cruciality at capsid-scale}
\label{sec:microscale}
As mentioned in Section \ref{sec:combinatorialEntropy}, given the free energy
and reaction rate of each subassembly of all the nodes and the structure of the
assembly tree, we can define its likelihood under the assumption of successful
assembly and we can group assembly trees into assembly pathways based on
some prediction-related criteria \cite{bona2011enumeration, bona2008influence,
sitharam2005counting}. To simplify our model, we abbreviate the notion of the
assembly pathway, to a \emph{connectivity pathway}, which only requires a test
of connectivity for the internal nodes of the assembly tree.

Informally, a \emph{connectivity pathway} corresponds directly to a minimal set
of interfaces that a successfully assembled capsid must contain to even be a
connected structure. It consists of the icosahedral symmetry classes of
assembly trees that use only this minimal set of interfaces, weighted by the
number of trees.

Given the dual graph $G=(V,E)$, of the capsid polyhedron (defined in Section
\ref{sec:nanoscale}), $E$ can be partitioned into sets $E_\iota$, one for each
interface type $\iota$ (for $T=1$ capsid polyhedra, there are 3 interface types
and for $T=3$ capsid polyhedra, there are 7 interface types as shown in
\figref{Fig 6}). A set $I$ of interface types is a \emph{connectivity
pathway} if the set of edges $E_I := \bigcup\limits_{\iota\in I} E_\iota$, is a
connected subgraph of $G$ and for each $\iota \in I$ $E_I \setminus E_\iota$ is
not connected.

\begin{figure}[htpb]
\centering
\begin{subfigure}{0.45\textwidth}
\includegraphics[scale=0.2]{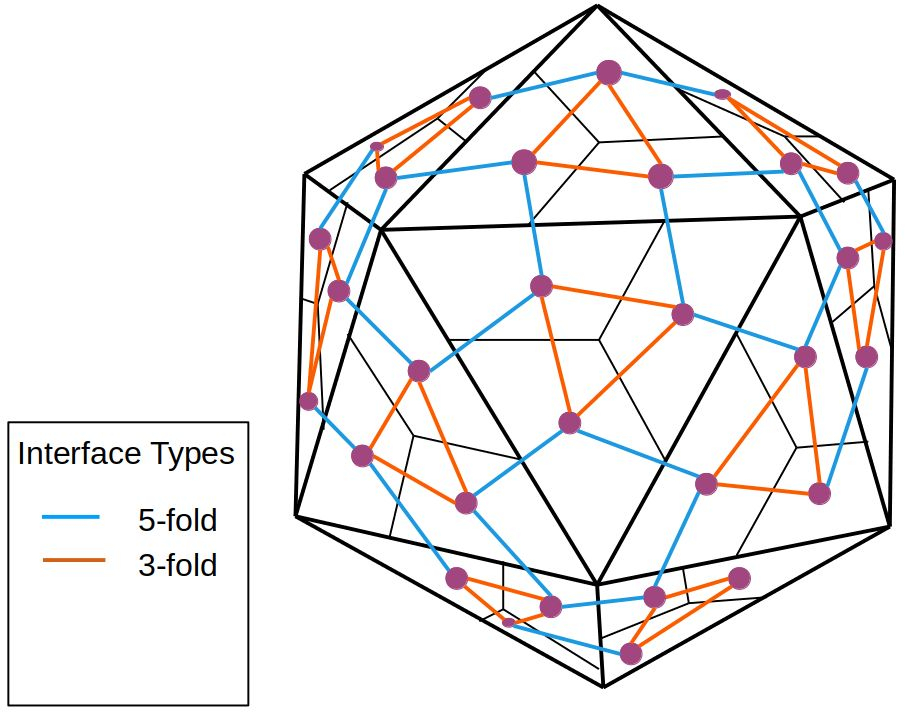}
\caption{\centering}
\label{fig:T1PentamerTrimer} 
\label{fig:T1All} 
\end{subfigure}
\begin{subfigure}{0.45\textwidth}
\includegraphics[scale=0.2]{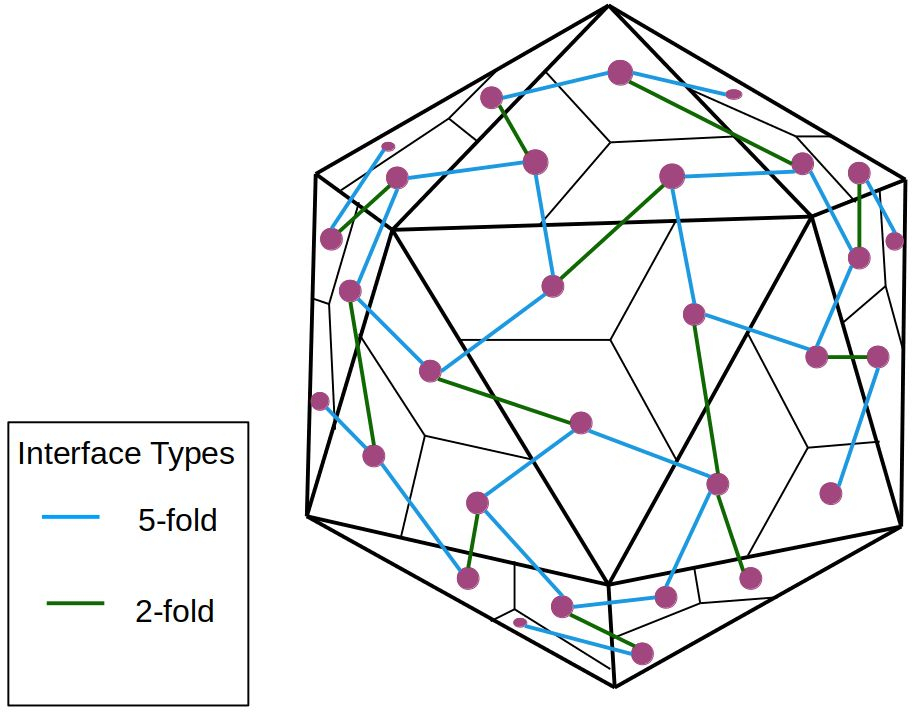}
\caption{\centering}
\label{fig:T1PentamerDimer} 
\end{subfigure}
\begin{subfigure}{0.45\textwidth}
\includegraphics[scale=0.2]{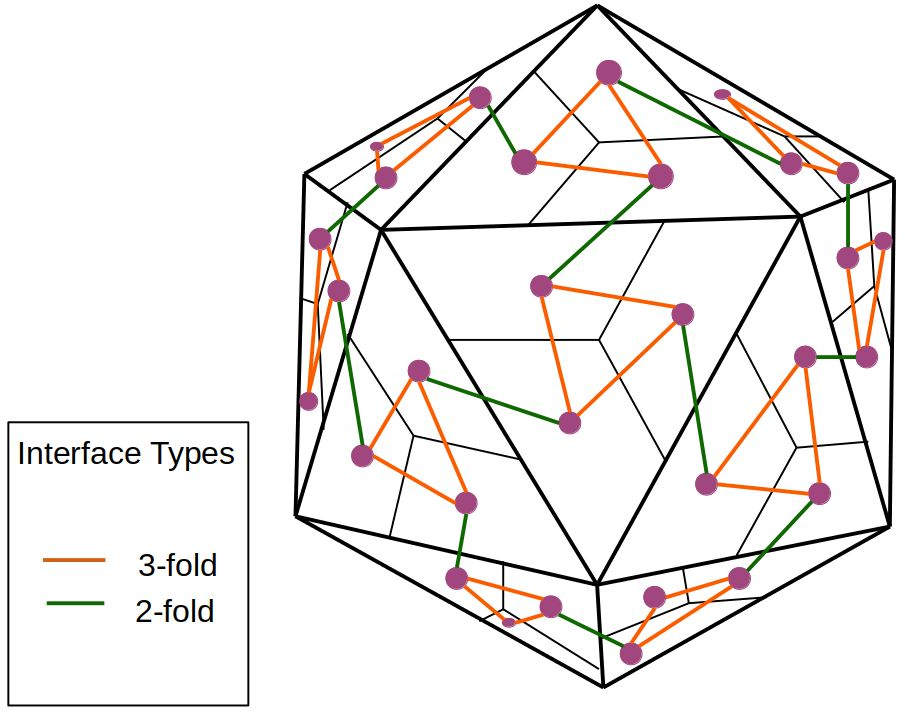}
\caption{\centering}
\label{fig:T1TrimerDimer}
\end{subfigure}
\caption{{\bf $T=1$ capsid polyhedra showing all 3 possible connectivity pathways.}
See Section \ref{sec:microscale}.}
\label{Fig 7}
\end{figure}

\begin{figure}[htpb]
\begin{subfigure}{0.45\textwidth}
\centering
\includegraphics[scale=0.24]{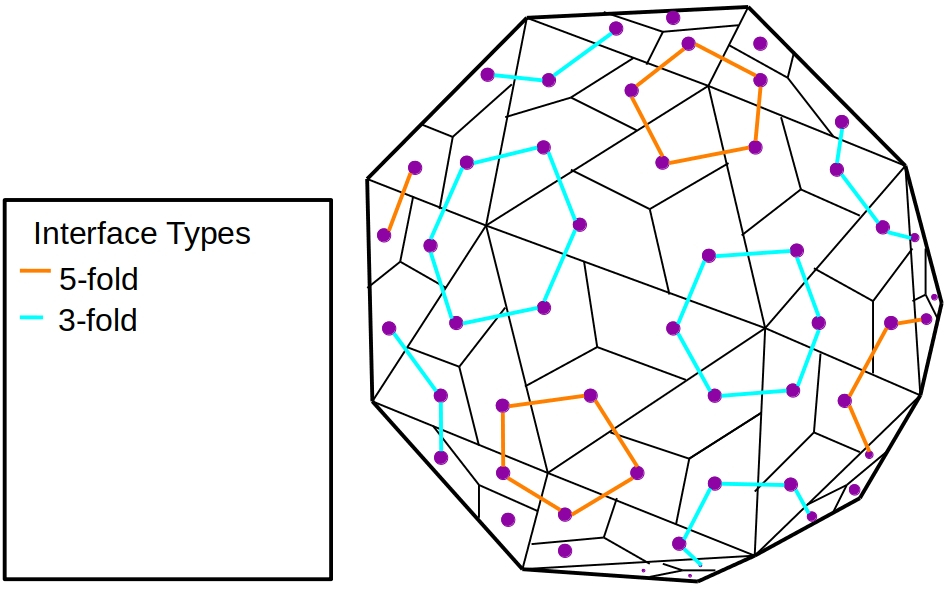}
\caption{\centering}
\label{fig:T=3PentamerHexamer} 
\end{subfigure}
\begin{subfigure}{0.45\textwidth}
\includegraphics[scale=0.22]{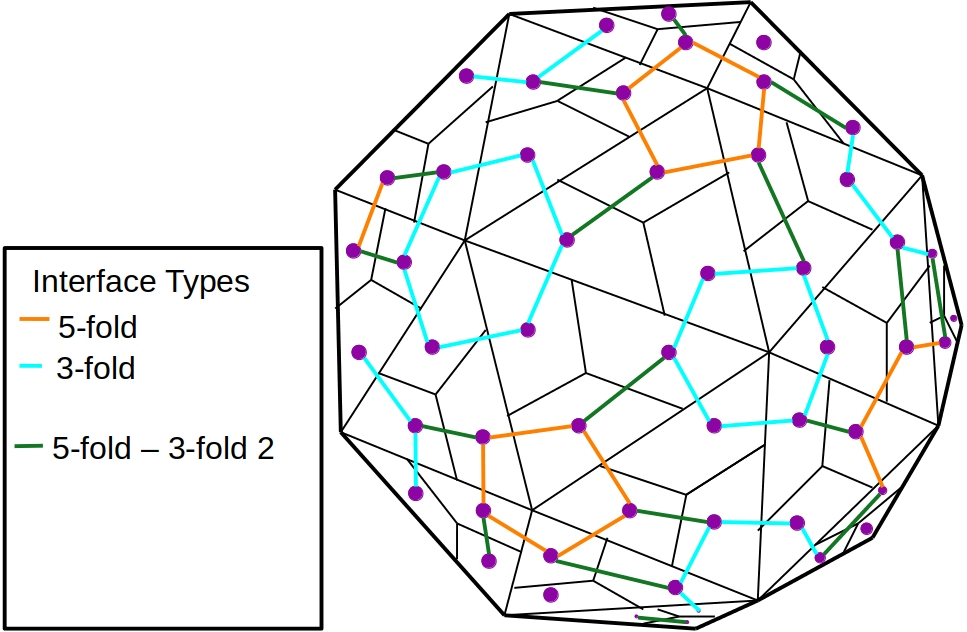}
\caption{\centering}
\label{fig:T=3PentamerHexamerPen-Hex}
\end{subfigure}
\caption{{\bf $T=3$ capsid polyhedra.} (a) Only the 5 fold and 3 fold interfaces are
shown. This does not correspond to a connectivity pathway. (b) 5 fold, 3 fold
and 5 fold-3 fold 2 interfaces are shown. This corresponds to a connectivity
pathway.  See Section \ref{sec:microscale}.}
\label{Fig 8}
\end{figure}

Given the small number of interface types for a capsid polyhedron of any $T$
number, we can find all connectivity pathways using a simple graph algorithm.
\figref{Fig 7}, shows 3 sets $I$ of interface types, for a $T=1$
capsid polyhedron that correspond to connectivity pathways.
\figref{Fig 8} shows 2 sets $I$ of interface types for a $T=3$ capsid polyhedron,
one of which is a connectivity pathway and one that is not.
The \emph{cruciality of an interface type} is the number of connectivity
pathways containing that interface type.

\subsection{Two-scale prediction: interaction cruciality at capsid-scale}
\label{sec:multiscale}
We use two different types of two-scale predictions. The first prediction
assumes that all interface types are equally important and is based only on the
cruciality of interactions to interface types. For an interface of type
$\iota$, the probability $P_\iota^r$ of breaking the interface when dropping an
interaction $r$ is measured by the the cruciality bar-code: $(\mu_{minima}^r,
\mu_{capsid}^r)$, as described in Section \ref{sec:nanoscale}. Results
validating these predictions are shown in Section
\ref{sec:results:ConfPrediction}.

For the second two-scale prediction, we combine the interaction cruciality at
the interface-scale and the interface cruciality at the capsid-scale using a
statistical model as follows.

A simple linear model is used to learn the relative weights $a_\iota$,
$b_\iota$, and $c_\iota$ for $$ P_\iota^r = \sigma(a_\iota \cdot \mu_{minima}^r
+ b_\iota \cdot \mu_{capsid}^r + c_\iota) $$ where $\sigma$ is the standard
sigmoid or threshold function used in neural networks. The training data are
obtained from site-directed mutagenesis results found in literature
\cite{wu2000mutational, okinaka2001c, bleker2005mutational, reguera2004role,
lochrie2006mutations, riolobos2006nuclear, UFDC000005707},
that measure the effect of removing an interaction $r$ on capsid assembly.

In addition, we learn the relative weights of the two scales through a scalar
parameter $w_\iota \in [0, 1]$ which represents the cruciality of an interface
type $\iota$ to any connectivity pathway. The probability of breaking a
connectivity pathway when an interaction is dropped is approximated by the
equation:
\begin{equation}
C^r_p = 1-\prod_{\iota \in p} (1- w_\iota \cdot P_\iota^r)
\label{eq:prob_break_path}
\end{equation}

For example, when $w_\iota = 0$, the corresponding term in
Eq. \eqref{eq:prob_break_path} vanishes, and breaking any interface of type
$\iota$ has no effect on disrupting assembly. Conversely, when all $w_\iota$
are equal, the probability of disruption depends only on $P^r_\iota$'s, namely
the cruciality of the interactions to interfaces and the number of connectivity
paths in which an interface participates.

Putting these together, we get the \emph{cruciality of an interaction $r$ for
capsid assembly} given by $$ H(r) = \sum_p C^r_p.$$ In this model, the
parameters $a_\iota$, $b_\iota$, $c_\iota$, and $w_\iota$ are all unknown. We
determine their value using simple machine learning. For a given partial order
over the interactions $T = \{(r_i, r_j): r_i$ has bigger impact on the capsid
assembly than $ r_j\}$, the cruciality function $H$ should satisfy $H(r_i) > H(r_j)$.
Towards this end, we design a loss function: $$ L = \sum_{(r_i, r_j) \in T}
\sigma (H(r_i) - H(r_j))$$ where $\sigma$ is the standard sigmoid or threshold
function used in neural networks. When the cruciality function $H$ satisfies
the partial order, the loss function will be minimal. So the parameters can be
determined by evaluating $$ \argmin_{a_\iota, b_\iota, c_\iota, w_\iota} L $$
Results validating these predictions are shown in Section
\ref{sec:results:ConvPrediction}.

\section{Results}
\label{sec:results}
The ab-initio computational predictions were blind to the results of the
directed mutagenesis biophysical assays, noting that the biophysical assays for
AAV2 were carried out by the Mavis Agbandje-Mckenna's lab \cite{UFDC000005707}
contemporaneously with development of our computational model and prediction.
Section \ref{sec:results:dataValidation} describes the in-vitro mutagenesis
setup. Section \ref{sec:predictionInput} describes the input and output of the
computational model used for prediction. Section
\ref{sec:results:ConfPrediction} describes the validation of the interaction
cruciality prediction assuming that all interface types are equally important
and is hence effectively based only on the interaction cruciality for interface
assembly. Section \ref{sec:results:ConvPrediction} describes the validation of
the two-scale prediction which combines the interaction cruciality at the
interface-scale and the interface cruciality at the capsid-scale using a
statistical model.

\subsection{Experimental setup} \label{sec:results:dataValidation}
We validate our prediction for AAV2 ($T=1$), MVM ($T=1$), and BMV ($T=3$) viral
capsids, using site-directed mutagenesis and biophysical assays to characterize
the variants generated. AAV2, MVM, and BMV residues were selected based on
their location in the 2-fold, 3-fold, and 5-fold symmetry-related interface of
the viral capsid, and alanine scanning of all charged residues in the VP. 
The mutagenesis results used for validation were found in literature
\cite{wu2000mutational, okinaka2001c, bleker2005mutational, reguera2004role,
lochrie2006mutations, riolobos2006nuclear, UFDC000005707}. We note
that the biophysical assays for AAV2 were carried out by the Mavis
Agbandje-Mckenna's lab \cite{UFDC000005707} contemporaneously with the
development of our computational model and prediction.

\eat{Plasmids containing the VP were mutated by site-directed mutagenesis. Mutant
plasmids were expressed in their permissive cell lines to generate the VP
variants. The variants generated were characterized by anti-capsid antibodies,
variants that did not assemble or produce significantly less capsids than the
wild type plasmid were defective for capsid assembly. To confirm the presence
of assembled viral capsids, the expressed variants were loaded on a
centrifugation gradient and further purified by column chromotography. The
virus containing fractions were concentrated and analyzed by coomassie stained
SDS PAGE and negative stain electron microscopy (EM).}

\begin{table}[htbp]
\centering
		\caption{Mutagesis data used in this manuscript for BMV, MVM, and AAV2
		along with the sources of the data.} 
\begin{subtable}[t]{0.48\textwidth}
\resizebox{\columnwidth}{!}{%
\begin{tabular}{|c|c|c|}
\hline
Residue & Mutagenesis result & Source \\ \hline
227 & Disrupt & \cite{wu2000mutational} \\ \hline
231 & Disrupt & \cite{wu2000mutational} \\ \hline
232 & Disrupt & \cite{wu2000mutational} \\ \hline
292 & Disrupt & \cite{wu2000mutational} \\ \hline
294 & Disrupt & \cite{UFDC000005707}\\ \hline
297 & Disrupt & \cite{UFDC000005707}\\ \hline
298 & Disrupt & \cite{UFDC000005707}\\ \hline
334 & Non-Disrupt & \cite{bleker2005mutational} \\ \hline
337 & Non-Disrupt & \cite{bleker2005mutational} \\ \hline
382 & Non-Disrupt & \cite{lochrie2006mutations} \\ \hline
389 & Non-Disrupt & \cite{UFDC000005707}\\ \hline
397 & Disrupt & \cite{UFDC000005707}\\ \hline
402 & Disrupt & \cite{UFDC000005707}\\ \hline
661 & Non-Disrupt & \cite{UFDC000005707}\\ \hline
692 & Disrupt & \cite{UFDC000005707}\\ \hline
694 & Disrupt & \cite{UFDC000005707}\\ \hline
696 & Disrupt & \cite{wu2000mutational} \\ \hline
704 & Non-Disrupt & \cite{lochrie2006mutations} \\ \hline
706 & Non-Disrupt & \cite{lochrie2006mutations} \\ \hline
\end{tabular}
}
\caption{\centering AAV2}
\end{subtable}
\begin{subtable}[t]{0.48\textwidth}
\begin{tabular}{|c|c|c|}
\hline
Residue & Mutagenesis result & Source \\ \hline
55 & Disrupt & \cite{reguera2004role} \\ \hline
129 & Disrupt & \cite{reguera2004role} \\ \hline
153 & Disrupt & \cite{reguera2004role} \\ \hline
168 & Non-Disrupt& \cite{reguera2004role} \\ \hline
261 & Non-Disrupt & \cite{UFDC000005707}\\ \hline
302 & Disrupt & \cite{riolobos2006nuclear} \\ \hline
507 & Non-Disrupt & \cite{UFDC000005707}\\ \hline
540 & Non-Disrupt& \cite{UFDC000005707}\\ \hline
543 & Disrupt & \cite{reguera2004role} \\ \hline
546 & Disrupt & \cite{reguera2004role} \\ \hline
567 & Disrupt & \cite{riolobos2006nuclear} \\ \hline
\end{tabular}
\caption{\centering MVM}
\end{subtable}
\begin{subtable}[t]{0.48\textwidth}
\begin{tabular}{|c|c|c|}
\hline
Residue & Mutagenesis result & Source \\ \hline
51 & Disrupt & \cite{okinaka2001c} \\ \hline
180 & Partial Disrupt & \cite{okinaka2001c} \\ \hline
181 & Disrupt & \cite{okinaka2001c} \\ \hline
182 & Disrupt & \cite{okinaka2001c} \\ \hline
183 & Partial Disrupt & \cite{okinaka2001c} \\ \hline
184 & Disrupt & \cite{okinaka2001c} \\ \hline
185 & Non-Disrupt & \cite{okinaka2001c} \\ \hline
188 & Partial Disrupt & \cite{okinaka2001c} \\ \hline
189 & Partial Disrupt & \cite{okinaka2001c} \\ \hline
\end{tabular}
\caption{\centering BMV}
\end{subtable}
\label{tab:mutagenesis}
\end{table}

The residues were classified by the yield of successfully assembled capsids
compared to wild type after the mutation: a yield of 100\% indicates that
mutation has no effect on the assembly and the residue is marked
\emph{non-disrupt}; a yield of 0\% indicates the assembly is completely
disrupted, and the residue is marked as \emph{disrupt}
\cite{bleker2005mutational, kern2003identification, lochrie2006mutations,
okinaka2001c, perez2011molecular, reguera2004role, riolobos2006nuclear,
sacher1989effects, wu2000mutational}. 

\begin{figure}[htpb]
\centering
\begin{subfigure}{0.45\textwidth}
\includegraphics[width=\textwidth]{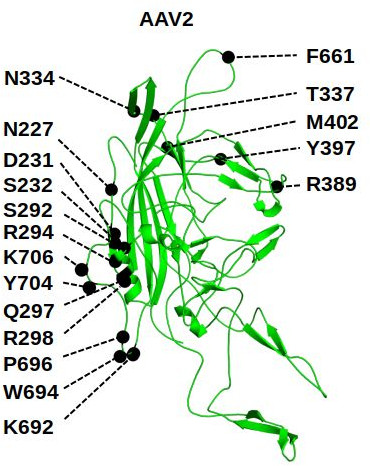}
\caption{\centering AAV2 monomer}
\label{fig:pbmv_chart} 
\end{subfigure}
\begin{subfigure}{0.45\textwidth}
\includegraphics[width=\textwidth]{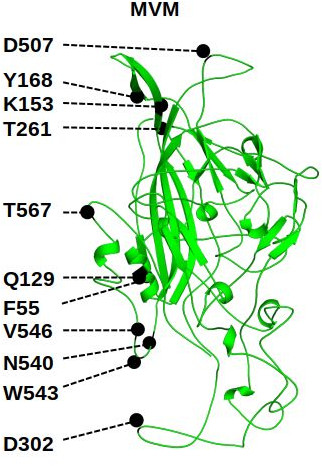}
\caption{\centering MVM monomer}
\label{fig:pbmv_chart} 
\end{subfigure}
\begin{subfigure}{0.45\textwidth}
\includegraphics[width=\textwidth]{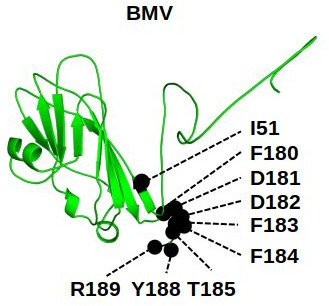}
\caption{\centering BMV monomer}
\label{fig:pbmv_chart}
\end{subfigure}
\caption{\bf AAV2\cite{Drouin8542}, MVM\cite{Llamas-Saiz:gr0639},
and BMV\cite{LUCAS200295} monomers showing the list of residues that 
were analyzed in this manuscript.}
\label{Fig 9}
\end{figure}

All the in-vitro mutagenesis results used in this manuscript are 
detailed in Table \ref{tab:mutagenesis}, along with the sources of the data. 
Figure \ref{Fig 9} shows these residues on a cartoon of
their respective VP monomers.

\subsection{Input and output of computational prediction model}
\label{sec:predictionInput}
For the interface-scale prediction, we started from simplified potential
energies designed from known X-ray structure of the VP monomers of each of the
viruses and all their interfaces \cite{Drouin8542, Llamas-Saiz:gr0639,
LUCAS200295} We treated the participating VP monomers or dimers as single rigid
motifs in the interface assembly systems.

The potential energy includes the hard-sphere potential between all atom pairs
(one from each participating VP monomer or dimer) with the Van der Waals radius
set to 1.2 \AA. We used Lennard-Jones pair potentials, setting the energy
difference of Eq. \eqref{eq:EnergyDiff} to $E_h - E_l =0.997 kJ/Mol$
\cite{Rahman1964}, and the weight in Eq. \eqref{eq:weights} to $w(N_a) \approx
1.5^{N_a}$. An implicit solvent was assumed.

Interface assembly landscapes were atlased using EASAL using the X-ray 3D
structures of the participating VP monomers and dimers and the above-described
pair potentials as input. Altogether 9 such atlases were obtained for different
interface assembly systems for each of the $T=1$ interface types and 31 such
atlases for the $T=3$ interface types, as described in Section
\ref{sec:nanoscale}. Each atlas computation takes no more than a couple of
hours on a laptop with Intel Core i5-2500K @ 3.2 Ghz CPU and 8GB of RAM.
Computations of cruciality required modification and analysis of each atlas for
each interaction (approximately 20 per interface).  Furthermore, we took into
account the simultaneous disabling of all interactions involving a residue, as
occurs in site-directed mutagenesis experiments. These analyses took
microseconds.

In all cases, the interface-scale predictions were performed blindly without
knowledge of in-vitro site directed mutagenesis results concerning
assembly-driving interactions. In particular, for AAV2, mutagenesis results
were only obtained subsequent to the interface-scale predictions. For MVM and
BMV, the in-vitro site directed mutagenesis results were also gathered,
subsequent to the interface-scale prediction, from multiple sources
\cite{bleker2005mutational, kern2003identification, lochrie2006mutations,
okinaka2001c, perez2011molecular, reguera2004role, riolobos2006nuclear,
sacher1989effects, wu2000mutational}. For the training phase of the second
two-scale prediction, less than half of the mutagenesis results were used,
picking pairs of interactions marked disrupt and non-disrupt. Both training and
learning phases took microseconds for each virus.

\subsubsection{Justifying input assumptions}
There are two aspects of the Lennard-Jones potential that could possibly
affect our prediction: (a) the energy difference $E_h - E_l$ and (b) the width
we set for the discretized Lennard-Jones well. Our prediction method relies
significantly more on the width of the discretized Lennard-Jones well rather
than the actual energy difference $E_h - E_l$. The number of atom-pairs within
the well essentially determines the quantity $N_a$ which is the deciding
quantity in the computation of the normalized partition function affecting the
cruciality prediction. However, since we know that the given capsid assembly
configuration is feasible, the minimum pairwise distance (which in all cases
turned out to be Van der Waals) forces the lower bound of the width of the
discretized Lennard-Jones well. On the other hand, since the given capsid
assembly configuration is a local energy minimum with (at least) 6 atom-pairs
within the Lennard-Jones well, we can measure a natural upper bound value for
the width of the discretized Lennard-Jones well (a larger value would permit a
energy-neutral motion within the basin to a configuration with more atom-pairs
in the Lennard-Jones well, and hence lower-energy, contradicting the minimality
of the given capsid assembly configuration.) 

Although, theoretically, the potential energy should include the Lennard-Jones
potential of all atom pairs, only the set of atom pairs that are close enough
to interact and are conserved in related viruses have noticeable contribution
to the configurational entropy. For the different types of interfaces (3 types
for $T=1$ and 7 types for $T=3$), we determined such pairs of interacting
residues (10-20 pairs for each interface), called the \emph{candidate
interactions} of each interface.

\subsubsection{Output of interface-based cruciality prediction}
\label{sec:results:ConfPrediction}
\begin{figure}[htpb]
\begin{subfigure}{0.45\textwidth}
\includegraphics[width=\linewidth]{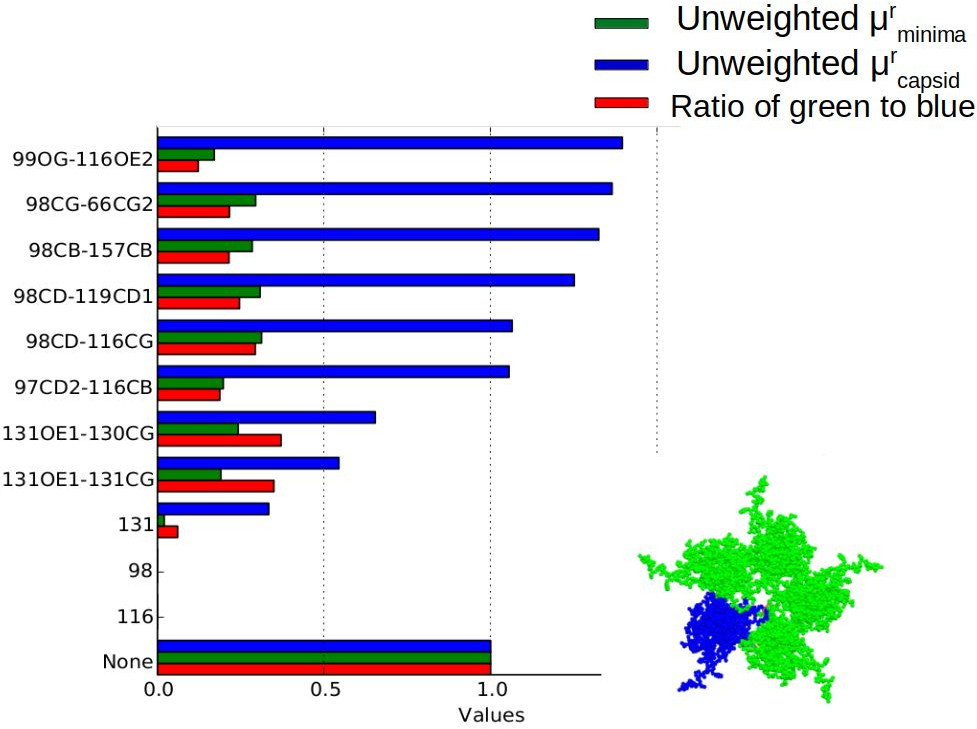}
\caption{\centering BMV 5-fold with VP monomer}
\label{fig:pbmv_chart} 
\end{subfigure}
\begin{subfigure}{0.45\textwidth}
\includegraphics[width=\linewidth]{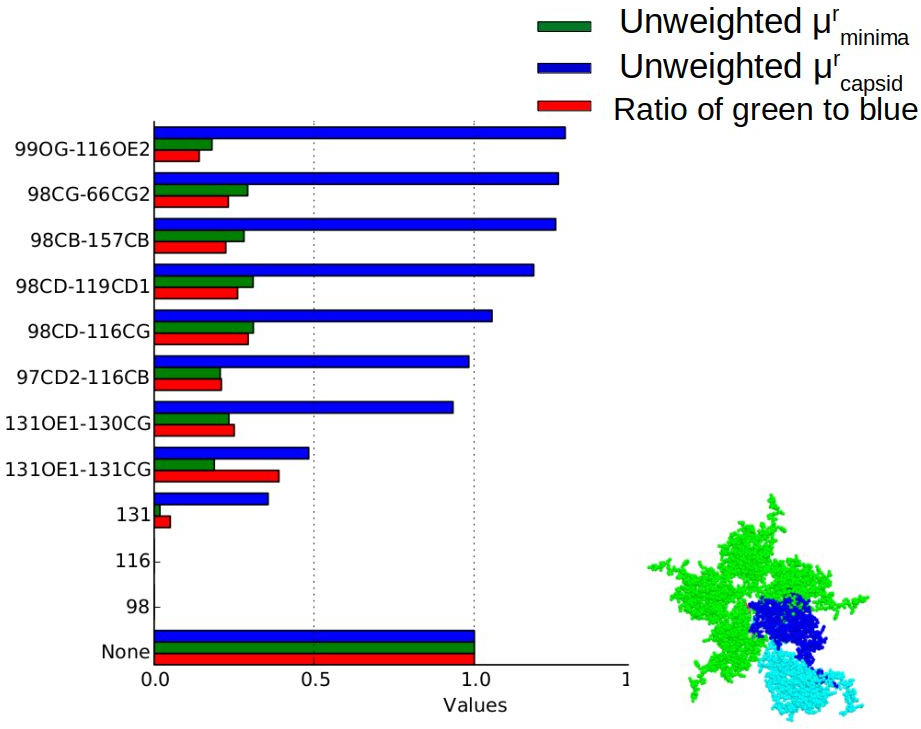}
\caption{\centering BMV 5-fold with VP dimer}
\label{fig:pbmv_d2_chart} 
\end{subfigure}
\caption{ {\bf Cruciality bar-codes.} Each row shows $\mu_{minima}^r$,
$\mu_{capsid}^r$ and their ratio in two BMV 5-fold interface assembly systems
(VP monomer-monomer and VP monomer-dimer shown at bottom right) where the
interaction $r$ - listed as the row label - is removed. The row labeled `None'
is the ``wild-type'' assembly system where no interaction has been removed,
whose $\nu_{minima}$ and $\nu_{capsid}$ values have been normalized. The rows
are sorted according to the largest value of $\mu_{capsid}$. See Section
\ref{sec:results:ConfPrediction}.}
\label{Fig 10}
\end{figure}

As discussed in Section \ref{sec:nanoscale}, for each interface assembly system
$s$, we use the unweighted versions of the cruciality bar-code
$(\mu^{r,s}_{minima}, \mu^{r,s}_{capsid})$ to predict the cruciality of an
interaction to an interface. \figref{Fig 10}(a) shows the plot of of the these
two parameters for the interface assembly system $s$ being the 5-fold interface
with VP monomers for BMV. Each row shows $\mu^{r,s}_{minima}$,
$\mu^{r,s}_{capsid}$ and their ratio in two BMV 5-fold interface assembly
systems (shown at the bottom right) where the interaction $r$ (which is the row
label) is removed. The row labeled `None' is the \emph{wild type} assembly
system where no interaction has been removed. The wild type system has been
used to normalize the values of all the other rows.  The rows are sorted
according to the largest value of $\mu^r_{capsid}$.

\figref{Fig 10}(b) plots the same parameters, but instead of considering VP
monomers assembling at the 5-fold interface, we consider the assembly of a VP
monomer and a VP dimer (as shown to the bottom right). As explained in Section
\ref{sec:nanoscale}, sterics play a larger role during the assembly of VP
dimers than during the assembly of VP monomers. Note that certain interactions
that had a lower value of $\mu_{capsid}$ in \figref{Fig 10}(a) have a higher
value in \figref{Fig 10}(b). Since these plots merely illustrate our
predictions without comparing them to mutagenesis data, the interested reader
is referred to the link in Section \ref{sec:suppInfo} for the complete set of
such data, for all the interface assembly systems for all the viruses.

\subsection{Validating the first two-scale prediction}
\label{sec:twoScale1}
\begin{figure}[htpb]
\centering
\begin{subfigure}{0.3\textwidth}
\includegraphics[width=\linewidth]{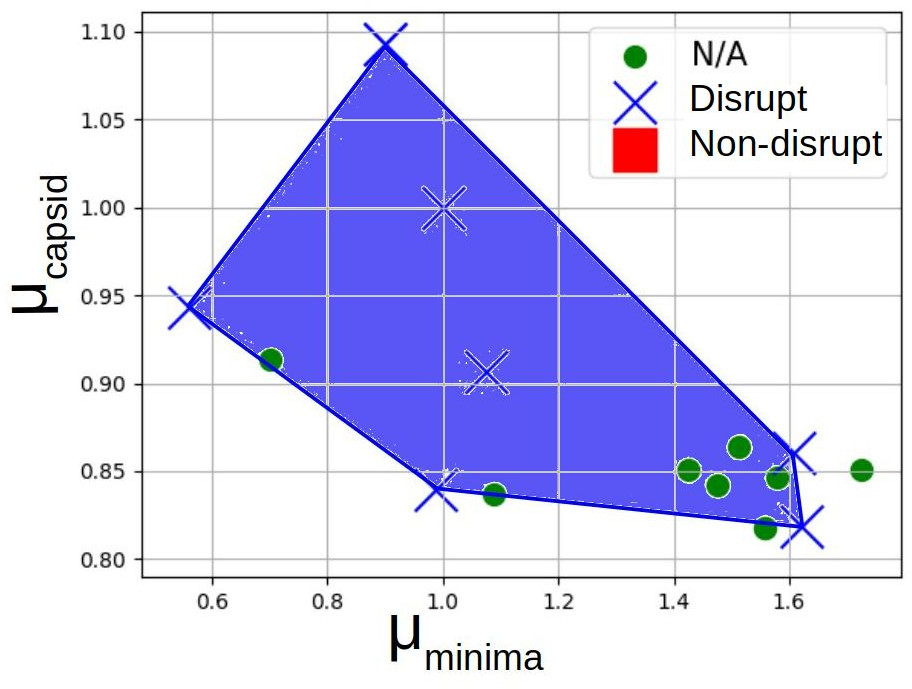}
\caption{\centering 2-fold interface in AAV2}
\label{fig:daav_init} 
\end{subfigure}
\begin{subfigure}{0.3\textwidth}
\centering
\includegraphics[width=\linewidth]{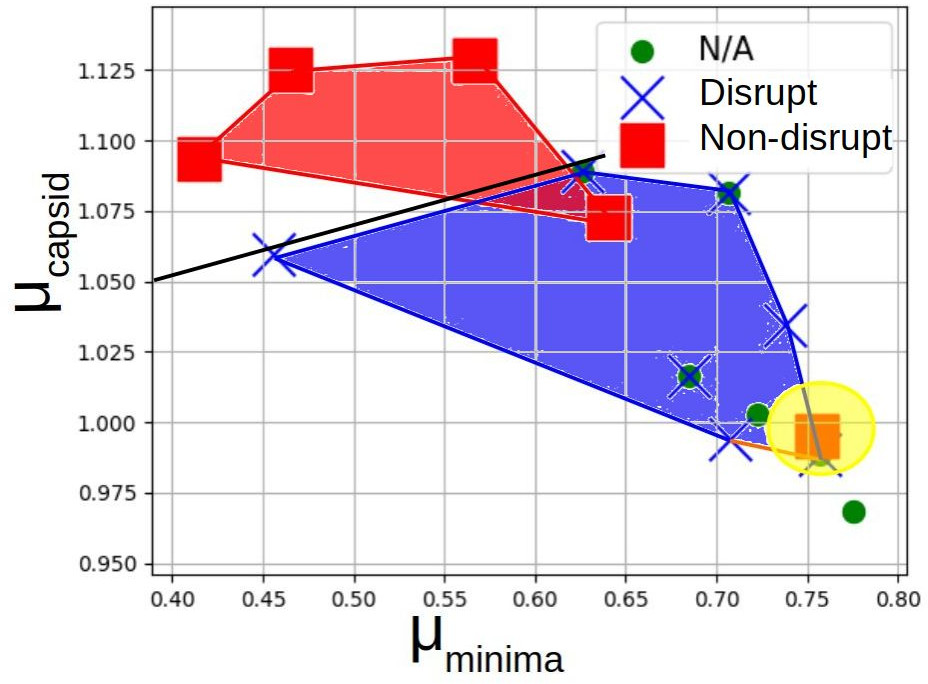}
\caption{\centering 5-fold interface in AAV2}
\label{fig:paav_init}	
\end{subfigure}
\begin{subfigure}{0.3\textwidth}
\centering
\includegraphics[width=\linewidth]{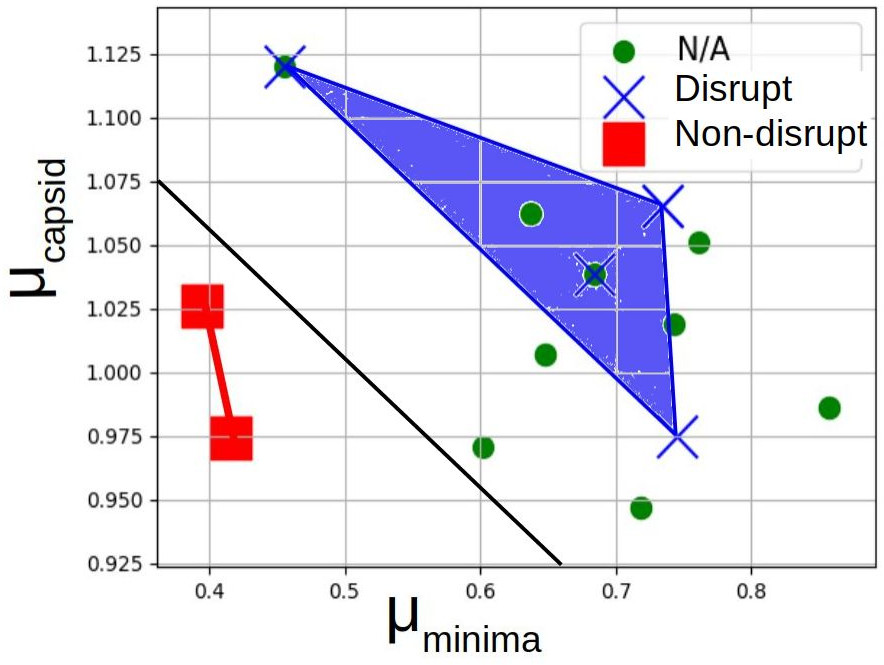}
\caption{\centering 2-fold interface in MVM}
\label{fig:pbmv_d2_chart} 
\end{subfigure}
\begin{subfigure}{0.3\textwidth}
\centering
\includegraphics[width=\linewidth]{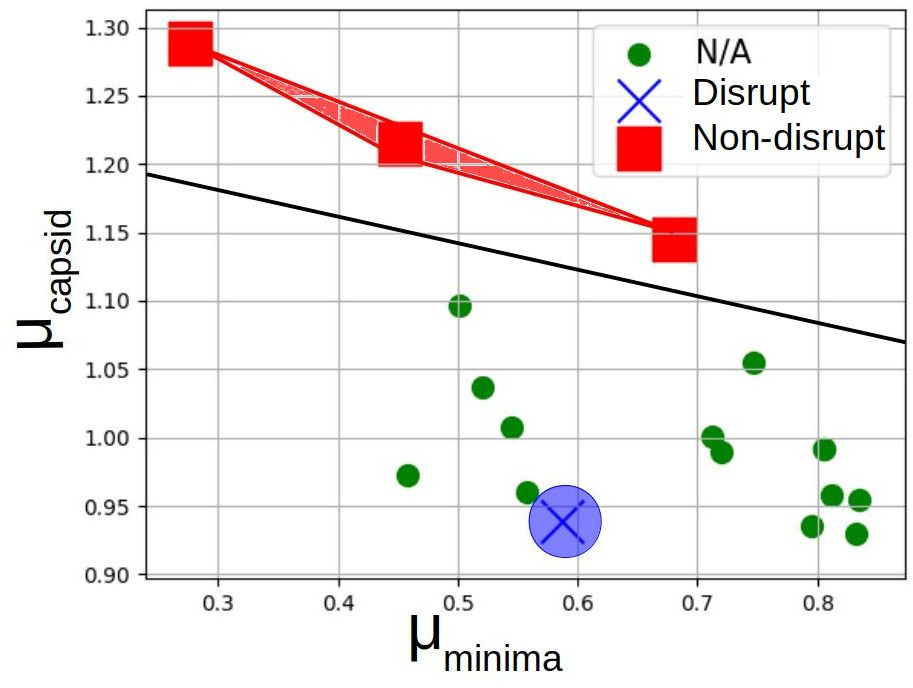}
\caption{\centering 5-fold interface in MVM}
\label{fig:pmvmp_init} 
\end{subfigure}
\begin{subfigure}{0.3\textwidth}
\centering
\includegraphics[width=\linewidth]{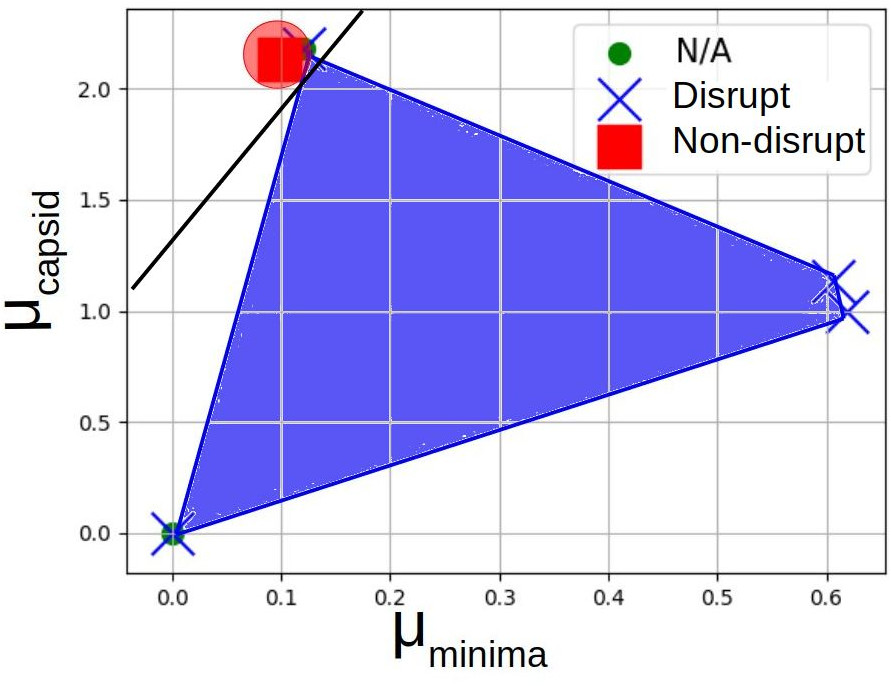}
\caption{\centering 2-fold interface in BMV}
\label{fig:dbmv_init}
\end{subfigure}
\begin{subfigure}{0.3\textwidth}
\centering
\includegraphics[width=\linewidth]{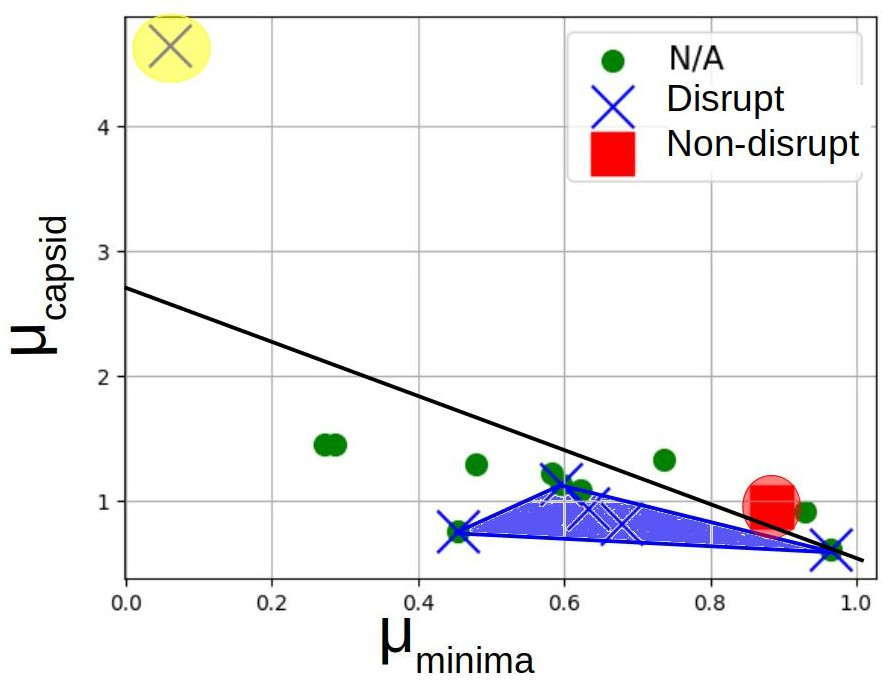}
\caption{\centering 5-fold - 3-fold interface in BMV}
\label{fig:d2bmv_init}
\end{subfigure}
\begin{subfigure}{0.3\textwidth}
\centering
\includegraphics[width=\linewidth]{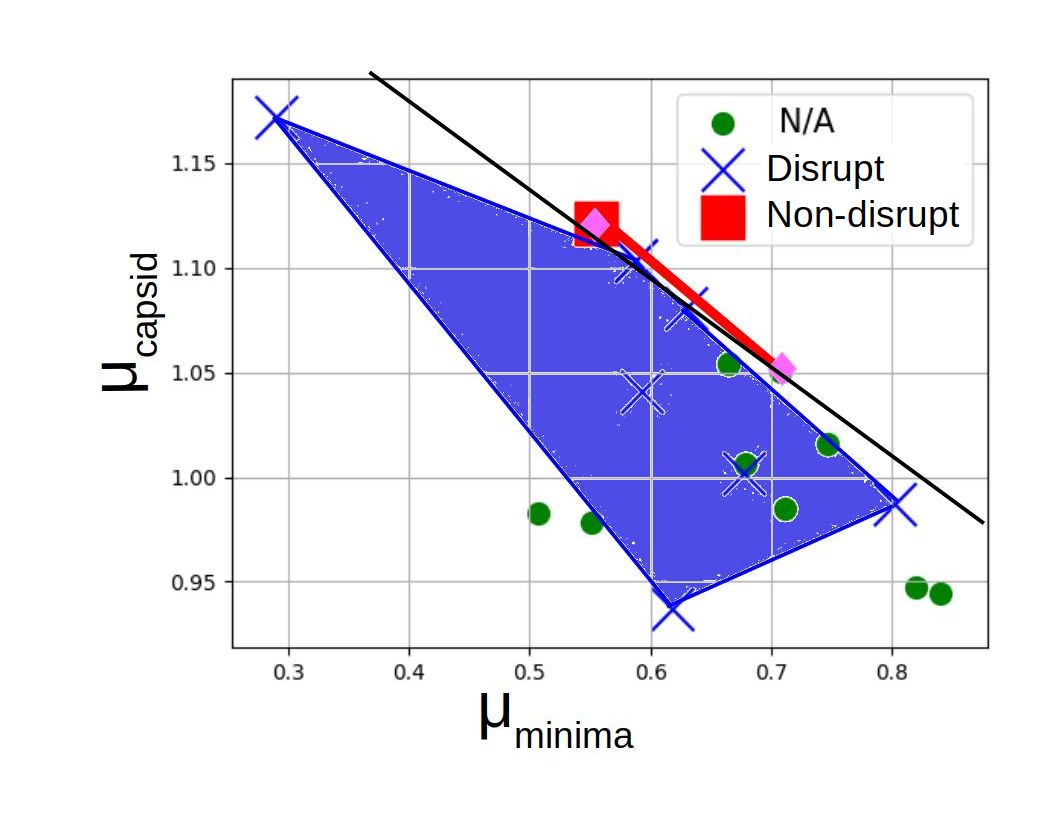}
\caption{\centering 2-fold interface in AAV2 with extra interactions}
\end{subfigure}
\begin{subfigure}{0.3\textwidth}
\centering
\includegraphics[width=\linewidth]{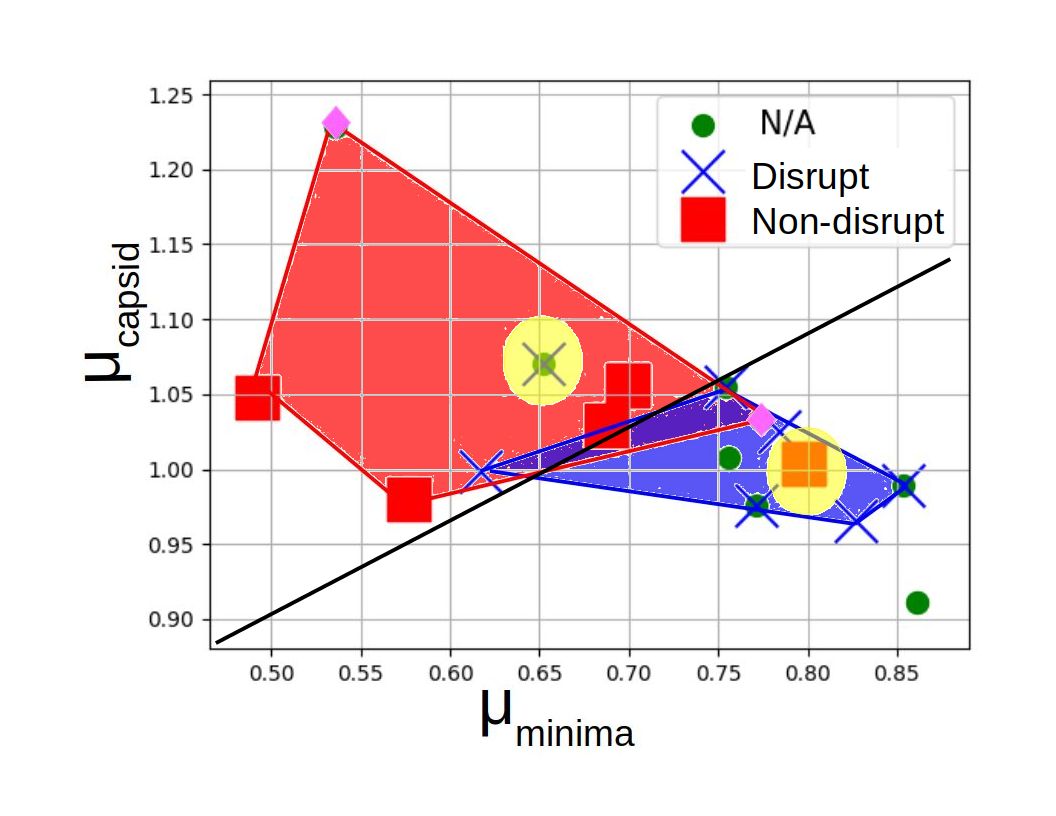}
\caption{\centering 5-fold interface in AAV2 with extra interactions}
\label{fig:paav_init_extra}
\end{subfigure}
\begin{subfigure}{0.3\textwidth}
\centering
\includegraphics[width=\linewidth]{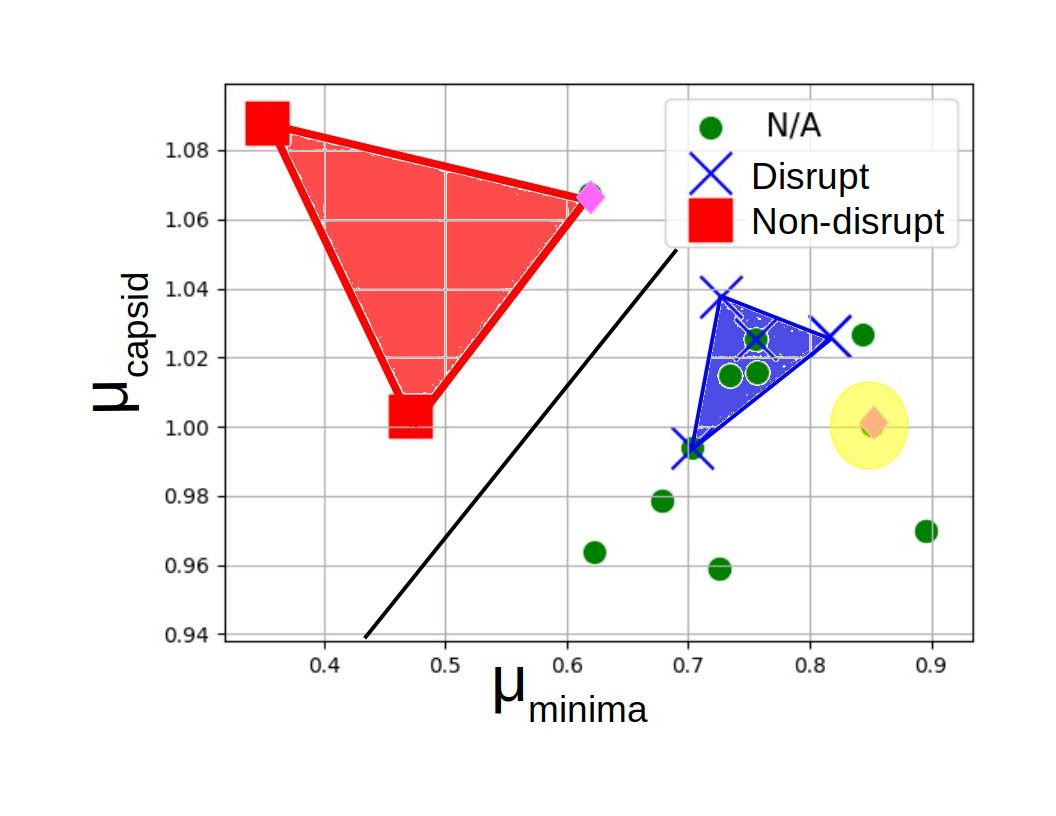}
\caption{\centering 2-fold interface in MVM with extra interactions}
\label{fig:dmvmp_init_extra}
\end{subfigure}
\begin{subfigure}{0.3\textwidth}
\centering
\includegraphics[width=\linewidth]{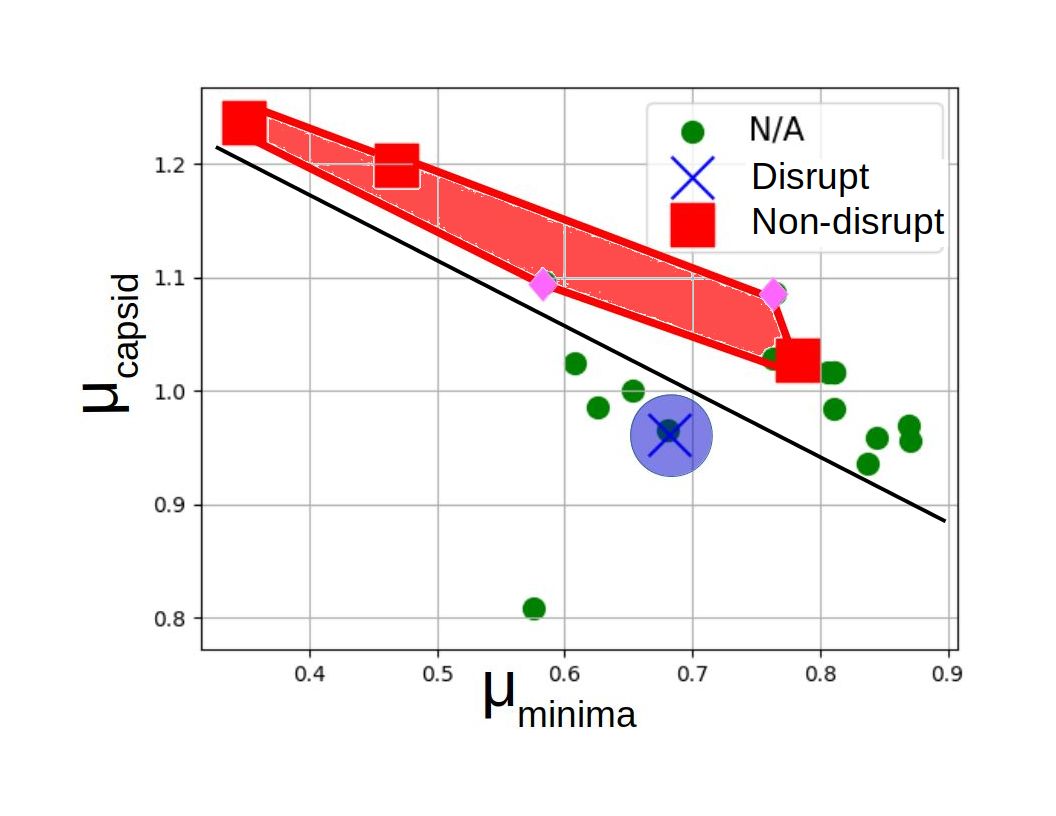}
\caption{\centering 5-fold interface in MVM with extra interactions}
\label{fig:pmvmp_init_extra} 
\end{subfigure}
\caption{{\bf Validation of direct interface-scale cruciality prediction: 2D plot of
cruciality bar-codes for each interface.} The blue cross marks and red squares
are residues found, through mutagenesis, to disrupt and non-disrupt assembly.
The green circles are residues on which no mutagenesis was performed. The
convex hulls are computational predictions from our method. The blue convex
hull delineates the residues that are shown to disrupt, the red convex hull
delineates the residues that are shown to not disrupt the assembly process,
yellow convex hull delineates the outliers.  In (g)-(j), the pink diamonds are
the extra interactions that were added to test for biases arising due to the
paucity of mutagenesis data.  The black line shows a linear separation of the
crucial and non-crucial residues.  See Section \ref{sec:twoScale1}.}
\label{Fig 11}
\end{figure}

\figref{Fig 11} shows the cruciality bar-code of residues for those interface
types (6 out of 13 interface types across the 3 viruses) for which there were
sufficient in-vitro site directed mutagenesis results for validation and for
which we were able to obtain cruciality predictions (see Section
\ref{sec:discussion}). As explained in Section \ref{sec:nanoscale}, the
cruciality bar-code for an interface type is a cumulative value obtained from
cruciality bar-codes computed for all the assembly systems at that interface.

Our interface-scale predictions were completely blind to the results of the
directed mutagenesis biophysical assays, noting that the biophysical assays for
AAV2 were carried out by the Mavis Agbandje-Mckenna's lab \cite{UFDC000005707}
contemporaneously with development of our computational model and prediction.
Although generalizing an interface-scale prediction to the capsid level assumes
the necessity of that interface for capsid assembly, our interface-scale
predictions were validated successfully using mutagenesis data towards capsid
assembly disruption. However, since this interface-scale prediction was part of
a second prediction (see Section \ref{sec:microscale}) using statistical
learning, that training data have been removed from \figref{Fig 11}.

The points in \figref{Fig 11} represent candidate hotspot residues.  The
coordinate values at which they have been placed, are our computational
cruciality predictions. The blue and red coloring of the points indicate
residues found through mutagenesis to disrupt and not disrupt assembly
respectively. The green circles are residues on which no mutagenesis was
performed. The blue convex hull delineates the residues that are shown to
disrupt, the red convex hull delineates the residues that are shown to not
disrupt. Yellow delineates outliers. The sub-figures show that the predicted
cruciality values of the residues that were later shown to disrupt assembly are
linearly separated from the predicted cruciality values of the residues that do
not disrupt assembly. I.e., the prediction convex hull formed by
assembly-disrupting residues does not significantly intersect the convex hull
formed by the non-assembly-disrupting residues. {\sl Conversely, if the
correlation between prediction and results were poor, there would not be such a
linear separation (or separation of convex hulls).}

For a reader interested in independently running the EASAL software to
reproduce our predictions, or in using other sources of experimental data to
check our predictions, we refer to the link in the Section \ref{sec:suppInfo}
containing a complete set of such cruciality bar-code plots, individually for
all the interface assembly systems, as well as the cumulative values for all
the interface types, for the 3 viruses.

To compensate for the paucity of in-vitro site directed mutagenesis results,
and to mitigate possible bias introduced when picking the candidate
interactions, we added 2 more candidate interactions to each interface. These
interactions are unlikely to be crucial, since they were not conserved across
similar viruses. Atlases were regenerated for each interface assembly system,
with these additional interactions and the cruciality bar-codes were computed
for all interactions using the new atlases. The results for the two $T=1$
viruses are shown as the last 4 figures of \figref{Fig 11}. Overall the added
residues (red convex hull) fall outside the blue convex hull delineating the
residues shown to disrupt assembly.

\subsection{Validating the second two-scale prediction}
\label{sec:results:ConvPrediction}
\begin{figure}[htpb]
\centering
\begin{subfigure}[t]{0.28\textwidth}
\centering
\includegraphics[width=\linewidth]{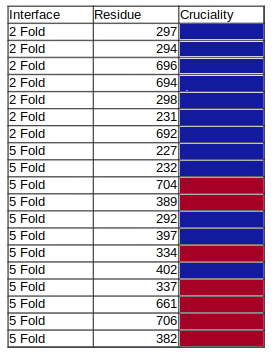}
\caption{\centering AAV2}
\label{fig:ResiduesAAV}
\end{subfigure}
\begin{subfigure}[t]{0.28\textwidth}
\centering
\includegraphics[width=\linewidth]{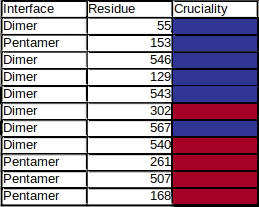}
\caption{\centering MVM}
\label{fig:ResiduesMVM}
\end{subfigure}
\begin{subfigure}[t]{0.3\textwidth}
\centering
\includegraphics[width=\linewidth]{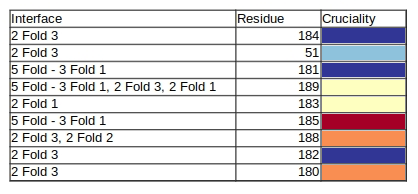}
\caption{\centering BMV}
\label{fig:ResiduesBMV} 
\end{subfigure}
\begin{subfigure}{0.1\textwidth}
\centering
\vspace{-1.8cm}
\label{fig:LearningLegend} 
\end{subfigure}
\caption{{\bf Validation of two-scale cruciality prediction using our statistical model 
for (a) AAV2, (b) MVM, and (c) BMV.} The residues listed higher
in the table are computationally predicted through our method as more 
crucial and the ones lower in the table are predicted as less crucial. 
Experimental mutagenesis results are used to mark all the residues by 
color. Blue indicates that the residue disrupts assembly while red 
indicates that it does not. See Section \ref{sec:results:ConvPrediction}.}
\label{Fig 12}
\end{figure}

\figref{Fig 12} shows, for AAV2, MVM and BMV, residues with their cruciality at
the capsid-scale, calculated using the statistical model of Section
\ref{sec:microscale}.  The correlation between our second two-scale prediction
and in vitro site directed mutagenesis data is illustrated by the correlation
between the prediction ranking of cruciality (top to bottom) and the
mutagenesis data shown (using in-vitro mutagenesis and biophysical assays)
extent of assembly disruption (color blue to red). More precisely, the residues
listed on the top of the table are computationally predicted through our method
to be more crucial, and the ones listed lower are predicted as less crucial. On
the other hand, residues shown (using in-vitro mutagenesis and biophysical
assays) to strongly disrupt assembly are blue and those that do not disrupt are
red, with partial disruption indicated by the spectrum of colors in between.
{\sl Conversely, if the correlation between prediction and mutagenesis data
were poor, the blues and reds would have been more interleaved.}

\begin{figure}[htpb]
\centering
\begin{subfigure}[t]{0.28\textwidth}
\centering
\includegraphics[width=\linewidth]{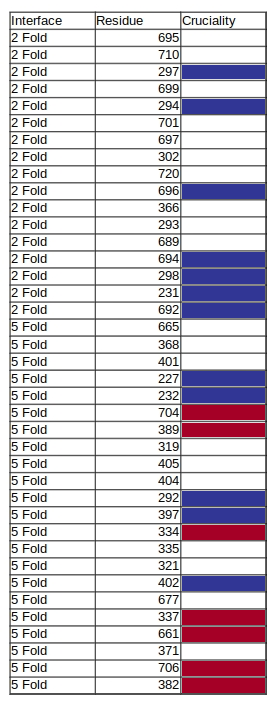}
\caption{\centering AAV2}
\label{fig:ResiduesAAV}
\end{subfigure}
\begin{subfigure}[t]{0.28\textwidth}
\centering
\includegraphics[width=\linewidth]{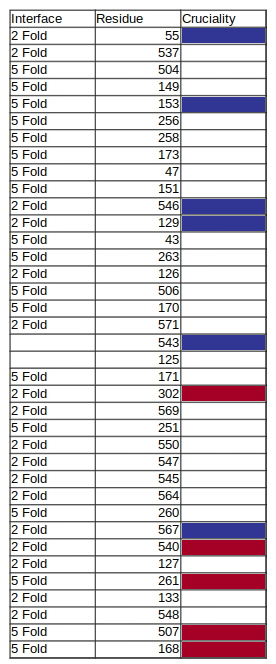}
\caption{\centering MVM}
\label{fig:ResiduesMVM}
\end{subfigure}
\begin{subfigure}[t]{0.3\textwidth}
\centering
\includegraphics[width=\linewidth]{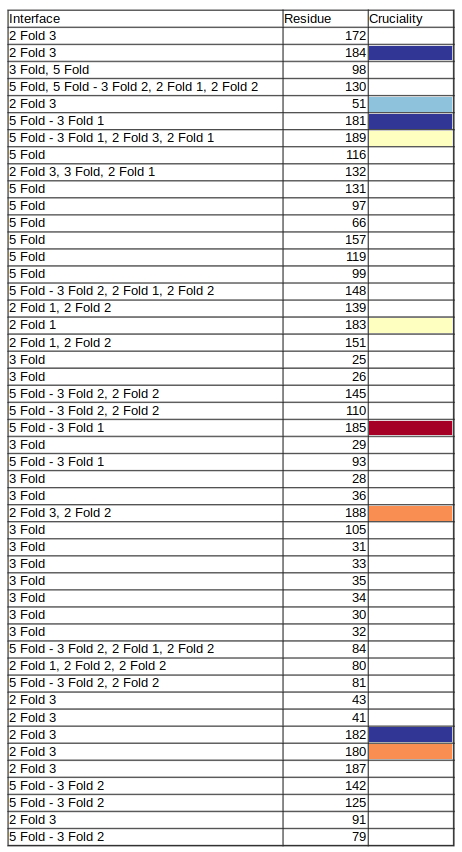}
\caption{\centering BMV}
\label{fig:ResiduesBMV} 
\end{subfigure}
\begin{subfigure}{0.1\textwidth}
\centering
\vspace{-1.8cm}
\includegraphics[width=\linewidth]{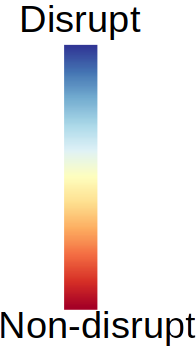}
\label{fig:LearningLegend} 
\end{subfigure}
\caption{{\bf Full list of two-scale cruciality predictions using our statistical
model for (a) AAV2, (b) MVM, and (c) BMV.} The residues listed higher in the
table are computationally predicted through our method as more crucial and the
ones lower in the table are predicted as less crucial.  Experimental
mutagenesis results are used to mark all the residues by color. Blue indicates
that the residue disrupts assembly while red indicates that it does not. White
indicates residues that do not yet have mutagenesis results for validation at
the time of this writing.  See Section \ref{sec:results:ConvPrediction}.}
\label{Fig 13}
\end{figure}

\figref{Fig 13} gives the full list of two-scale cruciality predictions using
our statistical model for the three viruses. As before, the residues listed on
the top of the table are computationally predicted through our method to be
more crucial, and the ones listed lower are predicted as less crucial. The
color codes have the same significance as explained in the previous paragraph.
In addition, white indicates residues that do not yet have mutagenesis results
for validation at the time of this writing.

As can be seen from \figref{Fig 13}, the amount of available
biophysical assays data for BMV was quite small when compared to
the other two viruses. This paucity leads to the BMV results not showing the
strong correlation seen in the results of the other two viruses. Despite this, it
still correctly predict the cruciality for most of the residues for which
in-vitro mutagenesis results are available.

\section{Discussion}
\label{sec:discussion}
Our prediction of crucial hotspot inter-atomic interactions between VP monomers
for the assembly of icosahedral viral capsids in 3 viruses, starts from a
candidate list of such interactions gleaned through sequence and structure
conservation in evolutionarily similar viruses. This data was provided by Dr.
Mavis Agbandje-Mckenna's lab. The crucial interaction prediction at the
interface-scale is purely using statistical mechanics: it is not
knowledge-based. We use an atlas (computed using the EASAL methodology) to
approximate the changes in the partition function of the capsid. The prediction
of interface cruciality at the capsid-scale uses an approximation of
combinatorial entropy. One of our two types of predictions uses statistical
learning to relatively weight the predictions at the two individual scales.
site-directed mutagenesis and biophysical assays results validate both types of
predictions.

Kinetics have clearly been shown to affect many aspects of viral capsid
assembly. While our method does not {\sl separately} model kinetics, it
combines interface-scale and capsid-scale analyses of free-energy and
configurational entropy, which could affect kinetics. Our predictions do
strongly rely on thermodynamics, specifically free energy. We believe that our
results show that geometry and consequently configurational entropy may already
be determinative in answering the narrow question of which inter-atomic
interactions are more or less crucial to assembly, at least for the viral
capsids studied here. 

Besides a planned comparison of our interface-scale crucial interaction
prediction method with a host of prevailing hotspot PPI prediction methods
\cite{Ibarra2019hotspot} on the SKEMPI database \cite{jankauskaite2019skempi},
there are several observations we made during the development of the method
that may lead to future work.

{\sl Additional interactions:} The candidate interactions that serve as the
input to EASAL are hand picked and pre-screened. Some interactions are excluded
because they are not likely to be crucial based some prior experience and some
are excluded since no mutation on that residue is possible for now. This could
potentially introduce bias in that the picked interaction are already likely
more crucial than the others. In addition, since other non-crucial interactions
also contribute to the potential energy, ignoring them will change the energy
landscape. Using an extended set of candidate interactions as input would
improve accuracy of prediction. Validation however would require more extensive
mutagenesis results.

{\sl Rigidity of the 3 and 5 fold interfaces:} As explained in Section
\ref{sec:backgroundEntropy}, we decompose the viral capsids into interface
\emph{bi-assembly} systems involving two assembling units. However, for the 3
and 5 fold interfaces, simultaneous tri-assembly and pent-assembly should be
considered. Better prediction could be obtained using newer variants of EASAL
that handle more than two input assembling units \cite{maineasal}.

{\sl Omitted interfaces:} Our results in Section
\ref{sec:results:ConfPrediction} do not show the predictions of interaction
cruciality for all interface types. Some of these omitted interfaces did not
have mutagenesis results for validation, and have been included in the link in
Section \ref{sec:suppInfo}. However, there are some interface that are not
shown in the supporting information as well, since we were unable to get any
useful predictions for these interfaces. These include the 3-fold interfaces
for $T=1$, $T=3$ virus and some 2-fold interfaces for $T=3$.

For 3-fold interfaces in AAV2, we could not obtain useful cruciality bar-codes
or rankings due to the heavy influence of sterics caused by interdigitation. In
addition, mutagenesis of the 3-fold interface interactions did not disrupt
assembly. We do not believe that the removal of any of the 3-fold interactions
causes assembly disruption. Most of the residues in the 3-fold interface of BMV
cross-link to the RNA and hence have no effect on assembly. We conjecture that
in these cases, the assembly proceeds primarily by 2-fold and 5-fold interface
interactions. Trimer interdigitation contributes to post-assembly stability of
the capsid.

{\sl BMV Results:} In the second two-scale prediction, the results for BMV do
not exhibit the strong correlation seen in the results of AAV2 and MVM. While
both AAV2 and MVM form empty capsid followed by the packing of the ssDNA
genome, BMV instead co-assembles the capsid with the ssRNA genome, which plays
an essential role in coordinating the assembly \cite{Comas_Garcia_2019}. While
empty BMV capsids can assemble in-vitro, the BMV wild-type has selected capsid
proteins that co-assemble with the genome. Since our methodology only considers
VP-monomers for crucial hotspot prediction, and doesn't take into account
co-assembly, the BMV results may be skewed. However, we beleive that the bigger
issue impacting the BMV results is the paucity of mutagenesis data available
for the training model. Ranking of residues which currently do no have
mutagenesis data for validation, are made available as part of the supporting
information (see Section \ref{sec:suppInfo}).

\section{Conclusion}
\label{sec:conclusion}
In this paper we predict crucial inter-atomic interactions between VP monomers
for the assembly of icosahedral viral capsids in 3 viruses, AAV2, MVM, and BMV.
The crucial interaction prediction at the interface-scale uses an atlas
generated with minimal sampling using the EASAL geometric methodology that
relies on convexifying landscape regions using Cayley parameters. From the
atlas, a cruciality bar-code approximates the changes in the partition function
of the capsid assembly landscape when an interaction is removed. At the
capsid-scale, an approximation of combinatorial entropy is used to predict the
cruciality of interface types at the capsid scale. We use 2 two-scale methods
to predict interface cruciality at the capsid scale. The first method is
entirely blind to known site-directed mutagenesis and biophysical assay
results, and assumes that each interface type is equally important for capsid
assembly and only uses interaction cruciality at the interface scale to predict
interaction cruciality at the capsid scale. The second method takes the
variation among interface types into account, using statistical learning to
relatively weight the predictions at the two scales. Site-directed mutagenesis
towards assembly disruption are used to validate our predictions. The method,
being computationally lightweight, rapid (100 to 1000 times faster than
prevailing methods \cite{maineasal,easalSoftware, ozkan2014fast}), rigorous,
and reliable, could be used to narrow down the field of candidate
assembly-driving interactions for in-vitro experiments, or even computationally
intensive in-silico experiments. For reproducibility, the reader can access and
run the EASAL source code \cite{easalSoftware} with the help of descriptive
papers \cite{maineasal, Ozkan:toms}, user guide \cite{easalUserGuide} and video
tutorial \cite{easalVideo}, as well as all of our raw prediction data for
cruciality bar-codes at URL
\url{https://geoplexity.bitbucket.io/virusSuppInfo.html}. This data includes
EASAL predictions that could not be validated with the mutagenesis data we had
access to, but could be checked against future mutagenesis experiments. At the
interface-scale, the method is general enough to apply to any assembly system,
in particular those that occur at various stages of the viral life-cycle, or
during the action of tests and drugs. As Figure \ref{Fig 2} shows, our
single-scale methods can be mixed and matched piecemeal with prevailing methods
to leverage complementary strengths (e.g., interface-scale hotspot predictions,
or sequence and structure conservation, or extensive in-silico validation). We
are unaware of any previous study that spans the range from multiscale
statistical mechanical predictions of crucial hotspot inter-atomic interactions
to validation using site-directed mutagenesis results, as opposed to
Computational Alanine Scanning (CAS) and other computationally intensive
Molecular Dynamics (MD) in-silico validations (explained in detail in Section
\ref{sec:Motivation}).

\section{Supporting information}
\label{sec:suppInfo}
Supporting information including raw prediction data for cruciality bar-codes,
and ranking of residues which currently do no have mutagenesis data for
validation, are available at
\url{https://geoplexity.bitbucket.io/virusSuppInfo.html}.

The above-mentioned results, as well as all the results in the paper can be
reproduced using the opensource EASAL software, curated by ACM TOMS in the
collected algorithms of the ACM \cite{Ozkan:toms}. Software freely available on
Bitbucket at \url{http://bitbucket.org/geoplexity/easal}. A user guide
(\url{https:
//bitbucket.org/geoplexity/easal/src/master/CompleteUserGuide.pdf}) and a video
illustrating the features of the software
\url{https://cise.ufl.edu/\~sitharam/EASALvideo.mpeg} are also provided.

\section*{Acknowledgments}
The authors thank Dr. Mavis Agbandje-Mckenna and Dr. Antonette Bennet for (1)
compiling the data from the sources \cite{wu2000mutational, okinaka2001c,
bleker2005mutational, reguera2004role, lochrie2006mutations,
riolobos2006nuclear, UFDC000005707} in Table \ref{tab:mutagenesis}, (2) help
with creating Figure \ref{Fig 9}, and (3) for providing us the candidate list
of interactions, which served as the input to our model for crucial hotspot
predictions (see Section \ref{sec:discussion}). We also thank them for
contemporaneously carrying out the biophysical assays for site-directed
mutagenesis on AAV2. Their results \cite{UFDC000005707} along with others
\cite{wu2000mutational, okinaka2001c, bleker2005mutational, reguera2004role,
lochrie2006mutations, riolobos2006nuclear} helped validate our predictions.

We thank an insightful comment by an anonymous reviewer about our BMV
predictions, now included in Section \ref{sec:discussion}. 

This research was supported in part by NSF Grants DMS-0714912, CCF-1117695,
DMS-1563234, and DMS-1564480.
\bibliography{virus}

\begin{thebibliography}{10}

\bibitem{caspar1962physical}
D.~L. Caspar and A.~Klug, ``Physical principles in the construction of regular
  viruses,'' in {\em Cold Spring Harbor symposia on quantitative biology},
  vol.~27, pp.~1--24, Cold Spring Harbor Laboratory Press, 1962.

\bibitem{UFDC000005707}
A.~Bennett, {\em Biophysical Characterization of the Assembly and Disassembly
  of the Single Stranded DNA Viruses}.
\newblock University of Florida Digital Collections., [Gainesville, Fla:
  University of Florida, 2009.
\newblock In the series University of Florida Digital Collections.

\bibitem{maineasal}
R.~Prabhu, M.~Sitharam, A.~Ozkan, and R.~Wu, ``Atlasing of assembly landscapes
  using distance geometry and graph rigidity,'' 2020.
\newblock arxiv:1203.3811.

\bibitem{Ozkan:toms}
A.~Ozkan, R.~Prabhu, T.~Baker, J.~Pence, J.~Peters, and M.~Sitharam,
  ``Algorithm 990: Efficient atlasing and search of configuration spaces of
  point-sets constrained by distance intervals,'' {\em ACM Trans. Math.
  Softw.}, vol.~44, pp.~48:1--48:30, July 2018.

\bibitem{schlicksup2018hepatitis}
C.~J. Schlicksup, J.~C.-Y. Wang, S.~Francis, B.~Venkatakrishnan, W.~W. Turner,
  M.~VanNieuwenhze, and A.~Zlotnick, ``Hepatitis b virus core protein
  allosteric modulators can distort and disrupt intact capsids,'' {\em Elife},
  vol.~7, p.~e31473, 2018.

\bibitem{GoodVirus}
M.~Mietzsch and M.~Agbandje-McKenna, ``The good that viruses do,'' {\em Annual
  Review of Virology}, vol.~4, no.~1, pp.~iii--v, 2017.
\newblock PMID: 28961414.

\bibitem{elrad2010encapsulation}
O.~M. Elrad and M.~F. Hagan, ``Encapsulation of a polymer by an icosahedral
  virus,'' {\em Physical biology}, vol.~7, no.~4, p.~045003, 2010.

\bibitem{reguera2019kinetics}
D.~Reguera, J.~Hern{\'a}ndez-Rojas, and J.~G. Llorente, ``Kinetics of empty
  viral capsid assembly in a minimal model,'' {\em Soft matter}, vol.~15,
  no.~36, pp.~7166--7172, 2019.

\bibitem{tuncbag2016potential}
N.~Tuncbag, A.~Gursoy, O.~Keskin, and R.~Nussinov, ``The potential impact of
  recent developments in three-dimensional quantitative interaction proteomics
  on structural biology,'' 2016.

\bibitem{luque2020cryo}
D.~Luque and J.~R. Cast{\'o}n, ``Cryo-electron microscopy for the study of
  virus assembly,'' {\em Nature Chemical Biology}, vol.~16, no.~3,
  pp.~231--239, 2020.

\bibitem{casini2004vitro}
G.~L. Casini, D.~Graham, D.~Heine, R.~L. Garcea, and D.~T. Wu, ``In vitro
  papillomavirus capsid assembly analyzed by light scattering,'' {\em
  Virology}, vol.~325, no.~2, pp.~320--327, 2004.

\bibitem{Ozkan2011}
A.~Ozkan and M.~Sitharam, ``Easal: Efficient atlasing, analysis and search of
  molecular assembly landscapes,'' in {\em Proceedings of the ISCA 3rd
  International Conference on Bioinformatics and Computational Biology},
  BICoB-2011, pp.~233--238, 2011.

\bibitem{easalSoftware}
A.~Ozkan, R.~Prabhu, T.~Baker, J.~Pence, and M.~Sitharam, ``Efficient atlasing
  and search of assembly landscapes ({ACM TOMS} version),'' 2016.

\bibitem{ozkan2014fast}
A.~Ozkan, J.~C. Flores-Canales, M.~Sitharam, and M.~Kurnikova, ``Fast and
  flexible geometric method for enhancing mc sampling of compact configurations
  for protein docking problem,'' 2014.

\bibitem{wu2000mutational}
P.~Wu, W.~Xiao, T.~Conlon, J.~Hughes, M.~Agbandje-McKenna, T.~Ferkol,
  T.~Flotte, and N.~Muzyczka, ``Mutational analysis of the adeno-associated
  virus type 2 (aav2) capsid gene and construction of aav2 vectors with altered
  tropism,'' {\em Journal of virology}, vol.~74, no.~18, pp.~8635--8647, 2000.

\bibitem{okinaka2001c}
Y.~Okinaka, K.~Mise, E.~Suzuki, T.~Okuno, and I.~Furusawa, ``The c terminus of
  brome mosaic virus coat protein controls viral cell-to-cell and long-distance
  movement,'' {\em Journal of virology}, vol.~75, no.~11, pp.~5385--5390, 2001.

\bibitem{bleker2005mutational}
S.~Bleker, F.~Sonntag, and J.~A. Kleinschmidt, ``Mutational analysis of narrow
  pores at the fivefold symmetry axes of adeno-associated virus type 2 capsids
  reveals a dual role in genome packaging and activation of phospholipase a2
  activity,'' {\em Journal of virology}, vol.~79, no.~4, pp.~2528--2540, 2005.

\bibitem{reguera2004role}
J.~Reguera, A.~Carreira, L.~Riolobos, J.~M. Almendral, and M.~G. Mateu, ``Role
  of interfacial amino acid residues in assembly, stability, and conformation
  of a spherical virus capsid,'' {\em Proceedings of the National Academy of
  Sciences}, vol.~101, no.~9, pp.~2724--2729, 2004.

\bibitem{lochrie2006mutations}
M.~A. Lochrie, G.~P. Tatsuno, B.~Christie, J.~W. McDonnell, S.~Zhou,
  R.~Surosky, G.~F. Pierce, and P.~Colosi, ``Mutations on the external surfaces
  of adeno-associated virus type 2 capsids that affect transduction and
  neutralization,'' {\em Journal of virology}, vol.~80, no.~2, pp.~821--834,
  2006.

\bibitem{riolobos2006nuclear}
L.~Riolobos, J.~Reguera, M.~G. Mateu, and J.~M. Almendral, ``Nuclear transport
  of trimeric assembly intermediates exerts a morphogenetic control on the
  icosahedral parvovirus capsid,'' {\em Journal of molecular biology},
  vol.~357, no.~3, pp.~1026--1038, 2006.

\bibitem{sitharam2006modeling}
M.~Sitharam and M.~Agbandje-Mckenna, ``Modeling virus self-assembly pathways:
  avoiding dynamics using geometric constraint decomposition,'' {\em Journal of
  Computational Biology}, vol.~13, no.~6, pp.~1232--1265, 2006.

\bibitem{sitharam:Assembly}
M.~Sitharam, {\em Modeling Autonomous Supramolecular Assembly}, pp.~197--216.
\newblock Berlin, Heidelberg: Springer Berlin Heidelberg, 2014.

\bibitem{sitharam2005counting}
M.~Sitharam and M.~B{\'o}na, ``Counting and enumeration of self-assembly
  pathways for symmetric macromolecular structures,'' in {\em Advances In
  Bioinformatics And Its Applications}, pp.~426--436, World Scientific, 2005.

\bibitem{bona2008influence}
M.~B{\'o}na and M.~Sitharam, ``The influence of symmetry on the probability of
  assembly pathways for icosahedral viral shells,'' {\em Computational and
  Mathematical Methods in Medicine}, vol.~9, no.~3-4, pp.~295--302, 2008.

\bibitem{bona2011enumeration}
M.~B{\'o}na, M.~Sitharam, and A.~Vince, ``Enumeration of viral capsid assembly
  pathways: Tree orbits under permutation group action,'' {\em Bulletin of
  mathematical biology}, vol.~73, no.~4, pp.~726--753, 2011.

\bibitem{Ibarra2019hotspot}
A.~A. Ibarra, G.~J. Bartlett, Z.~Heged{\"u}s, S.~Dutt, F.~Hobor, K.~A. Horner,
  K.~Hetherington, K.~Spence, A.~Nelson, T.~A. Edwards, D.~N. Woolfson, R.~B.
  Sessions, and A.~J. Wilson, ``Predicting and experimentally validating
  hot-spot residues at protein-protein interfaces,'' {\em ACS Chemical
  Biology}, vol.~14, pp.~2252--2263, Oct 2019.

\bibitem{rajamani2004anchor}
D.~Rajamani, S.~Thiel, S.~Vajda, and C.~J. Camacho, ``Anchor residues in
  protein--protein interactions,'' {\em Proceedings of the National Academy of
  Sciences}, vol.~101, no.~31, pp.~11287--11292, 2004.

\bibitem{xia2010apis}
J.-F. Xia, X.-M. Zhao, J.~Song, and D.-S. Huang, ``Apis: accurate prediction of
  hot spots in protein interfaces by combining protrusion index with solvent
  accessibility,'' {\em BMC bioinformatics}, vol.~11, no.~1, p.~174, 2010.

\bibitem{steinbrecher2017predicting}
T.~Steinbrecher, C.~Zhu, L.~Wang, R.~Abel, C.~Negron, D.~Pearlman, E.~Feyfant,
  J.~Duan, and W.~Sherman, ``Predicting the effect of amino acid single-point
  mutations on protein stability—large-scale validation of md-based relative
  free energy calculations,'' {\em Journal of molecular biology}, vol.~429,
  no.~7, pp.~948--963, 2017.

\bibitem{darnell2007decision}
S.~J. Darnell, D.~Page, and J.~C. Mitchell, ``An automated decision-tree
  approach to predicting protein interaction hot spots,'' {\em Proteins:
  Structure, Function, and Bioinformatics}, vol.~68, no.~4, pp.~813--823, 2007.

\bibitem{zhu2011kfc2}
X.~Zhu and J.~C. Mitchell, ``Kfc2: a knowledge-based hot spot prediction method
  based on interface solvation, atomic density, and plasticity features,'' {\em
  Proteins: Structure, Function, and Bioinformatics}, vol.~79, no.~9,
  pp.~2671--2683, 2011.

\bibitem{wang2012prediction}
L.~Wang, Z.-P. Liu, X.-S. Zhang, and L.~Chen, ``Prediction of hot spots in
  protein interfaces using a random forest model with hybrid features,'' {\em
  Protein Engineering, Design \& Selection}, vol.~25, no.~3, pp.~119--126,
  2012.

\bibitem{wang2014prediction}
L.~Wang, W.~Zhang, Q.~Gao, and C.~Xiong, ``Prediction of hot spots in protein
  interfaces using extreme learning machines with the information of spatial
  neighbour residues,'' {\em IET systems biology}, vol.~8, no.~4, pp.~184--190,
  2014.

\bibitem{ye2014prediction}
L.~Ye, Q.~Kuang, L.~Jiang, J.~Luo, Y.~Jiang, Z.~Ding, Y.~Li, and M.~Li,
  ``Prediction of hot spots residues in protein--protein interface using
  network feature and microenvironment feature,'' {\em Chemometrics and
  Intelligent Laboratory Systems}, vol.~131, pp.~16--21, 2014.

\bibitem{deng2014predhs}
L.~Deng, Q.~C. Zhang, Z.~Chen, Y.~Meng, J.~Guan, and S.~Zhou, ``Predhs: a web
  server for predicting protein--protein interaction hot spots by using
  structural neighborhood properties,'' {\em Nucleic acids research}, vol.~42,
  no.~W1, pp.~W290--W295, 2014.

\bibitem{sukhwal2015ppcheck}
A.~Sukhwal and R.~Sowdhamini, ``Ppcheck: A webserver for the quantitative
  analysis of protein-protein interfaces and prediction of residue hotspots,''
  {\em Bioinformatics and biology insights}, vol.~9, pp.~BBI--S25928, 2015.

\bibitem{sun2016accurate}
Z.~Sun, J.~Zhang, C.-H. Zheng, B.~Wang, and P.~Chen, ``Accurate prediction of
  protein hot spots residues based on gentle adaboost algorithm,'' in {\em
  International Conference on Intelligent Computing}, pp.~742--749, Springer,
  2016.

\bibitem{hu2017protein}
S.-S. Hu, P.~Chen, B.~Wang, and J.~Li, ``Protein binding hot spots prediction
  from sequence only by a new ensemble learning method,'' {\em Amino acids},
  vol.~49, no.~10, pp.~1773--1785, 2017.

\bibitem{murakami2017network}
Y.~Murakami, L.~P. Tripathi, P.~Prathipati, and K.~Mizuguchi, ``Network
  analysis and in silico prediction of protein--protein interactions with
  applications in drug discovery,'' {\em Current opinion in structural
  biology}, vol.~44, pp.~134--142, 2017.

\bibitem{Wang2018}
H.~Wang, C.~Liu, and L.~Deng, ``Enhanced prediction of hot spots at
  protein-protein interfaces using extreme gradient boosting,'' {\em Scientific
  Reports}, vol.~8, no.~1, p.~14285, 2018.

\bibitem{barradas2018structural}
D.~Barradas-Bautista, M.~Rosell, C.~Pallara, and J.~Fern{\'a}ndez-Recio,
  ``Structural prediction of protein--protein interactions by docking:
  Application to biomedical problems,'' in {\em Advances in protein chemistry
  and structural biology}, vol.~110, pp.~203--249, Elsevier, 2018.

\bibitem{liu2018machine}
S.~Liu, C.~Liu, and L.~Deng, ``Machine learning approaches for protein--protein
  interaction hot spot prediction: Progress and comparative assessment,'' {\em
  Molecules}, vol.~23, no.~10, p.~2535, 2018.

\bibitem{jankauskaite2019skempi}
J.~Jankauskait{\.e}, B.~Jim{\'e}nez-Garc{\'\i}a, J.~Dapk{\=u}nas,
  J.~Fern{\'a}ndez-Recio, and I.~H. Moal, ``Skempi 2.0: an updated benchmark of
  changes in protein--protein binding energy, kinetics and thermodynamics upon
  mutation,'' {\em Bioinformatics}, vol.~35, no.~3, pp.~462--469, 2019.

\bibitem{Diaz-Valle2019hotspot}
A.~D{\'\i}az-Valle, J.~M. Falc{\'o}n-Gonz{\'a}lez, and M.~Carrillo-Tripp, ``Hot
  spots and their contribution to the self-assembly of the viral capsid: In
  silico prediction and analysis,'' {\em International Journal of Molecular
  Sciences}, vol.~20, p.~5966, Nov 2019.

\bibitem{kaku}
M.~Karplus and J.~N. Kushick, ``Method for estimating the configurational
  entropy of macromolecules,'' {\em Macromolecules}, vol.~14, no.~2,
  pp.~325--332, 1981.

\bibitem{head1997mining}
M.~S. Head, J.~A. Given, and M.~K. Gilson, ``“mining minima”: direct
  computation of conformational free energy,'' {\em The Journal of Physical
  Chemistry A}, vol.~101, no.~8, pp.~1609--1618, 1997.

\bibitem{rapaport1999supramolecular}
D.~Rapaport, J.~Johnson, and J.~Skolnick, ``Supramolecular self-assembly:
  molecular dynamics modeling of polyhedral shell formation,'' {\em Computer
  physics communications}, vol.~121, pp.~231--235, 1999.

\bibitem{reddy1998energetics}
V.~S. Reddy, H.~A. Giesing, R.~T. Morton, A.~Kumar, C.~B. Post, C.~L.
  Brooks~III, and J.~E. Johnson, ``Energetics of quasiequivalence:
  computational analysis of protein-protein interactions in icosahedral
  viruses,'' {\em Biophysical journal}, vol.~74, no.~1, pp.~546--558, 1998.

\bibitem{andricioaei2001calculation}
I.~Andricioaei and M.~Karplus, ``On the calculation of entropy from covariance
  matrices of the atomic fluctuations,'' {\em The Journal of Chemical Physics},
  vol.~115, no.~14, pp.~6289--6292, 2001.

\bibitem{HAGAN200642}
M.~F. Hagan and D.~Chandler, ``Dynamic pathways for viral capsid assembly,''
  {\em Biophysical Journal}, vol.~91, no.~1, pp.~42 -- 54, 2006.

\bibitem{gfeller2007uncovering}
D.~Gfeller, D.~M. De~Lachapelle, P.~De~Los~Rios, G.~Caldarelli, and F.~Rao,
  ``Uncovering the topology of configuration space networks,'' {\em Physical
  Review E}, vol.~76, no.~2, p.~026113, 2007.

\bibitem{hnizdo2007nearest}
V.~Hnizdo, E.~Darian, A.~Fedorowicz, E.~Demchuk, S.~Li, and H.~Singh,
  ``Nearest-neighbor nonparametric method for estimating the configurational
  entropy of complex molecules,'' {\em Journal of computational chemistry},
  vol.~28, no.~3, pp.~655--668, 2007.

\bibitem{killian2007extraction}
B.~J. Killian, J.~Yundenfreund~Kravitz, and M.~K. Gilson, ``Extraction of
  configurational entropy from molecular simulations via an expansion
  approximation,'' {\em The Journal of chemical physics}, vol.~127, no.~2,
  p.~024107, 2007.

\bibitem{hnizdo2008efficient}
V.~Hnizdo, J.~Tan, B.~J. Killian, and M.~K. Gilson, ``Efficient calculation of
  configurational entropy from molecular simulations by combining the
  mutual-information expansion and nearest-neighbor methods,'' {\em Journal of
  computational chemistry}, vol.~29, no.~10, pp.~1605--1614, 2008.

\bibitem{Zhou_Gilson_2009}
H.-X. Zhou and M.~K. Gilson, ``Theory of free energy and entropy in noncovalent
  binding,'' {\em Chemical Reviews}, vol.~109, no.~9, pp.~4092--4107, 2009.
\newblock PMID: 19588959.

\bibitem{king2012efficient}
B.~M. King, N.~W. Silver, and B.~Tidor, ``Efficient calculation of molecular
  configurational entropies using an information theoretic approximation,''
  {\em The Journal of Physical Chemistry B}, vol.~116, no.~9, pp.~2891--2904,
  2012.

\bibitem{hensen2010estimating}
U.~Hensen, O.~F. Lange, and H.~Grubm{\"u}ller, ``Estimating absolute
  configurational entropies of macromolecules: The minimally coupled subspace
  approach,'' {\em PloS one}, vol.~5, no.~2, 2010.

\bibitem{fogolari2015distance}
F.~Fogolari, A.~Corazza, S.~Fortuna, M.~A. Soler, B.~VanSchouwen,
  G.~Brancolini, S.~Corni, G.~Melacini, and G.~Esposito, ``Distance-based
  configurational entropy of proteins from molecular dynamics simulations,''
  {\em PLoS One}, vol.~10, no.~7, 2015.

\bibitem{huang2011free}
D.~Huang and A.~Caflisch, ``The free energy landscape of small molecule
  unbinding,'' {\em PLoS computational biology}, vol.~7, no.~2, 2011.

\bibitem{dunton2014free}
T.~A. Dunton, J.~E. Goose, D.~J. Gavaghan, M.~S. Sansom, and J.~M. Osborne,
  ``The free energy landscape of dimerization of a membrane protein, nanc,''
  {\em PLoS computational biology}, vol.~10, no.~1, 2014.

\bibitem{staneva2011binding}
I.~Staneva and S.~Wallin, ``Binding free energy landscape of domain-peptide
  interactions,'' {\em PLoS computational biology}, vol.~7, no.~8, 2011.

\bibitem{prada2009exploring}
D.~Prada-Gracia, J.~G{\'o}mez-Garde{\~n}es, P.~Echenique, and F.~Falo,
  ``Exploring the free energy landscape: from dynamics to networks and back,''
  {\em PLoS computational biology}, vol.~5, no.~6, 2009.

\bibitem{varadhan2006topology}
G.~Varadhan, Y.~J. Kim, S.~Krishnan, and D.~Manocha, ``Topology preserving
  approximation of free configuration space,'' in {\em Proceedings 2006 IEEE
  International Conference on Robotics and Automation, 2006. ICRA 2006.},
  pp.~3041--3048, IEEE, 2006.

\bibitem{lai2009uncovering}
Z.~Lai, J.~Su, W.~Chen, and C.~Wang, ``Uncovering the properties of
  energy-weighted conformation space networks with a hydrophobic-hydrophilic
  model,'' {\em International journal of molecular sciences}, vol.~10, no.~4,
  pp.~1808--1823, 2009.

\bibitem{Chirikjian201199}
G.~S. Chirikjian, ``Chapter four - modeling loop entropy,'' in {\em Computer
  Methods, Part C} (M.~L. Johnson and L.~Brand, eds.), vol.~487 of {\em Methods
  in Enzymology}, pp.~99 -- 132, Academic Press, 2011.

\bibitem{Perilla2016}
J.~R. Perilla, J.~A. Hadden, B.~C. Goh, C.~G. Mayne, and K.~Schulten,
  ``All-atom molecular dynamics of virus capsids as drug targets,'' {\em The
  Journal of Physical Chemistry Letters}, vol.~7, pp.~1836--1844, May 2016.

\bibitem{Durrant2020}
J.~D. Durrant, S.~E. Kochanek, L.~Casalino, P.~U. Ieong, A.~C. Dommer, and
  R.~E. Amaro, ``Mesoscale all-atom influenza virus simulations suggest new
  substrate binding mechanism,'' {\em ACS Central Science}, vol.~6,
  pp.~189--196, Feb 2020.

\bibitem{statisticalMechanics2015}
S.~Whitelam and R.~L. Jack, ``The statistical mechanics of dynamic pathways to
  self-assembly,'' {\em Annual Review of Physical Chemistry}, vol.~66, no.~1,
  pp.~143--163, 2015.
\newblock PMID: 25493714.

\bibitem{virusMechansim2015}
J.~D. Perlmutter and M.~F. Hagan, ``Mechanisms of virus assembly,'' {\em Annual
  Review of Physical Chemistry}, vol.~66, no.~1, pp.~217--239, 2015.
\newblock PMID: 25532951.

\bibitem{zlotnick1994build}
A.~Zlotnick, ``To build a virus capsid: an equilibrium model of the self
  assembly of polyhedral protein complexes,'' {\em Journal of molecular
  biology}, vol.~241, no.~1, pp.~59--67, 1994.

\bibitem{zlotnick1999theoretical}
A.~Zlotnick, J.~M. Johnson, P.~W. Wingfield, S.~J. Stahl, and D.~Endres, ``A
  theoretical model successfully identifies features of hepatitis b virus
  capsid assembly,'' {\em Biochemistry}, vol.~38, no.~44, pp.~14644--14652,
  1999.

\bibitem{zlotnick2000mechanism}
A.~Zlotnick, R.~Aldrich, J.~M. Johnson, P.~Ceres, and M.~J. Young, ``Mechanism
  of capsid assembly for an icosahedral plant virus,'' {\em Virology},
  vol.~277, no.~2, pp.~450--456, 2000.

\bibitem{endres2002model}
D.~Endres and A.~Zlotnick, ``Model-based analysis of assembly kinetics for
  virus capsids or other spherical polymers,'' {\em Biophysical journal},
  vol.~83, no.~2, pp.~1217--1230, 2002.

\bibitem{ZLOTNICK2003536}
A.~Zlotnick and S.~J. Stray, ``How does your virus grow? understanding and
  interfering with virus assembly,'' {\em Trends in Biotechnology}, vol.~21,
  no.~12, pp.~536 -- 542, 2003.

\bibitem{zlotnick2005theoretical}
A.~Zlotnick, ``Theoretical aspects of virus capsid assembly,'' {\em Journal of
  Molecular Recognition: An Interdisciplinary Journal}, vol.~18, no.~6,
  pp.~479--490, 2005.

\bibitem{Hagan2010}
M.~F. Hagan and O.~M. Elrad, ``Understanding the concentration dependence of
  viral capsid assembly kinetics{\&}{\#}x2014;the origin of the lag time and
  identifying the critical nucleus size,'' {\em Biophysical Journal}, vol.~98,
  pp.~1065--1074, Mar 2010.

\bibitem{Bajaj-Quantified}
N.~Clement, M.~Rasheed, and C.~L. Bajaj, ``Viral capsid assembly: A quantified
  uncertainty approach,'' {\em Journal of Computational Biology}, vol.~25,
  no.~1, pp.~51--71, 2018.
\newblock PMID: 29313735.

\bibitem{schwartz}
R.~Schwartz, P.~W. Shor, P.~E. Prevelige~Jr, and B.~Berger, ``Local rules
  simulation of the kinetics of virus capsid self-assembly,'' {\em Biophysical
  journal}, vol.~75, no.~6, pp.~2626--2636, 1998.

\bibitem{berger1994local}
B.~Berger, P.~W. Shor, L.~Tucker-Kellogg, and J.~King, ``Local rule-based
  theory of virus shell assembly,'' {\em Proceedings of the National Academy of
  Sciences}, vol.~91, no.~16, pp.~7732--7736, 1994.

\bibitem{berger1994mathematics}
B.~Berger and P.~W. Shor, {\em The Mathematics of Virus Shell Assembly}.
\newblock MIT Center for Advanced Education Services, 1994.

\bibitem{berger1995local}
B.~Berger and P.~W. Shor, ``Local rules switching mechanism for viral shell
  geometry,'' in {\em Proc. 14th Biennial Conference on Phage Virus Assembly},
  Citeseer, 1995.

\bibitem{schwartz1998local}
R.~Schwartz, P.~E. Prevelige~Jr, and B.~Berger, ``Local rules modeling of
  nucleation-limited virus capsid assembly,'' {\em Massachusetts Institute of
  Technology, Cambridge, MA}, 1998.

\bibitem{rapaport2004self}
D.~Rapaport, ``Self-assembly of polyhedral shells: a molecular dynamics
  study,'' {\em Physical Review E}, vol.~70, no.~5, p.~051905, 2004.

\bibitem{rapaport2008role}
D.~Rapaport, ``Role of reversibility in viral capsid growth: a paradigm for
  self-assembly,'' {\em Physical Review Letters}, vol.~101, no.~18, p.~186101,
  2008.

\bibitem{polles2013mechanical}
G.~Polles, G.~Indelicato, R.~Potestio, P.~Cermelli, R.~Twarock, and
  C.~Micheletti, ``Mechanical and assembly units of viral capsids identified
  via quasi-rigid domain decomposition,'' {\em PLoS computational biology},
  vol.~9, no.~11, 2013.

\bibitem{pandey2014self}
S.~Pandey, D.~Johnson, R.~Kaplan, J.~Klobusicky, G.~Menon, and D.~H. Gracias,
  ``Self-assembly of mesoscale isomers: the role of pathways and degrees of
  freedom,'' {\em PloS one}, vol.~9, no.~10, 2014.

\bibitem{SJS:Handbook}
M.~Sitharam, A.~St.~John, and J.~Sidman, {\em Handbook of Geometric Constraint
  Systems Principles}.
\newblock CRC Press, 1st~ed., 2018.

\bibitem{kern2003identification}
A.~Kern, K.~Schmidt, C.~Leder, O.~M{\"u}ller, C.~Wobus, K.~Bettinger,
  C.~Von~der Lieth, J.~King, and J.~Kleinschmidt, ``Identification of a
  heparin-binding motif on adeno-associated virus type 2 capsids,'' {\em
  Journal of virology}, vol.~77, no.~20, pp.~11072--11081, 2003.

\bibitem{perez2011molecular}
R.~P{\'e}rez, M.~Castellanos, A.~Rodr{\'\i}guez-Huete, and M.~G. Mateu,
  ``Molecular determinants of self-association and rearrangement of a trimeric
  intermediate during the assembly of a parvovirus capsid,'' {\em Journal of
  molecular biology}, vol.~413, no.~1, pp.~32--40, 2011.

\bibitem{sacher1989effects}
R.~Sacher and P.~Ahlquist, ``Effects of deletions in the n-terminal basic arm
  of brome mosaic virus coat protein on rna packaging and systemic
  infection.,'' {\em Journal of virology}, vol.~63, no.~11, pp.~4545--4552,
  1989.

\bibitem{Drouin8542}
L.~M. Drouin, B.~Lins, M.~Janssen, A.~Bennett, P.~Chipman, R.~McKenna, W.~Chen,
  N.~Muzyczka, G.~Cardone, T.~S. Baker, and M.~Agbandje-McKenna,
  ``Cryo-electron microscopy reconstruction and stability studies of the wild
  type and the r432a variant of adeno-associated virus type 2 reveal that
  capsid structural stability is a major factor in genome packaging,'' {\em
  Journal of Virology}, vol.~90, no.~19, pp.~8542--8551, 2016.

\bibitem{Llamas-Saiz:gr0639}
A.~L. Llamas-Saiz, M.~Agbandje-McKenna, W.~R. Wikoff, J.~Bratton,
  P.~Tattersall, and M.~G. Rossmann, ``{Structure Determination of Minute Virus
  of Mice},'' {\em Acta Crystallographica Section D}, vol.~53, pp.~93--102, Jan
  1997.

\bibitem{LUCAS200295}
R.~W. Lucas, S.~B. Larson, and A.~McPherson, ``The crystallographic structure
  of brome mosaic virus,'' {\em Journal of Molecular Biology}, vol.~317, no.~1,
  pp.~95 -- 108, 2002.

\bibitem{Rahman1964}
A.~Rahman, ``Correlations in the motion of atoms in liquid argon,'' {\em Phys.
  Rev.}, vol.~136, pp.~A405--A411, Oct 1964.

\bibitem{Comas_Garcia_2019}
M.~Comas-Garcia, ``Packaging of genomic rna in positive-sense single-stranded
  rna viruses: A complex story,'' {\em Viruses}, vol.~11, p.~253, Mar 2019.

\bibitem{easalUserGuide}
R.~Prabhu and M.~Sitharam, ``{EASAL} software user guide.,'' 2016.

\bibitem{easalVideo}
R.~Prabhu, T.~Baker, and M.~Sitharam, ``Video illustrating the opensource
  software {EASAL},'' 2016.

\end{thebibliography}

\end{document}